\documentclass[a4paper,fleqn,usenatbib]{mnras}

\usepackage{mathptmx}

\usepackage[T1]{fontenc}
\usepackage{ae,aecompl}

\usepackage{graphicx}	
\usepackage{amsmath}	
\usepackage{amssymb}	
\usepackage{float}
\usepackage{subfigure}
\usepackage{color}
\usepackage{mathtools}
\usepackage{hhline}
\usepackage{multicol}
\usepackage{hyperref}

\usepackage{etoolbox}
\makeatletter
\patchcmd\@combinedblfloats{\box\@outputbox}{\unvbox\@outputbox}{}{%
   \errmessage{\noexpand\@combinedblfloats could not be patched}%
}%
 \makeatother

\title[Galaxy samples in photo-z surveys]{Optimizing galaxy samples for clustering measurements in photometric surveys}

\author[D. Tanoglidis et al.]{
Dimitrios Tanoglidis,$^{1,2}$\thanks{E-mail: dtanoglidis@uchicago.edu}
Chihway Chang,$^{1,2}$
and Joshua Frieman$^{1,2,3}$
\\
$^{1}$Department of Astronomy and Astrophysics, University of Chicago, Chicago, IL 60637, USA \\
$^{2}$Kavli Institute for Cosmological Physics, University of Chicago, Chicago, IL 60637, USA\\
$^{3}$Fermi National Accelerator Laboratory, P. O. Box 500, Batavia, IL 60510, USA
}

\date{Accepted XXX. Received YYY; in original form ZZZ}

\pubyear{2017}

\begin{document}
\label{firstpage}
\pagerange{\pageref{firstpage}--\pageref{lastpage}}
\maketitle

\begin{abstract}
When analyzing galaxy clustering in multi-band imaging surveys, there is a trade-off between selecting the largest galaxy samples (to minimize the shot noise) and selecting samples with the best photometric redshift (photo-z) precision, which generally include only a small subset of galaxies. In this paper, we systematically explore this trade-off. Our analysis is targeted towards the third year data of the Dark Energy Survey (DES), but our methods hold generally for other data sets. Using a simple Gaussian model for the redshift uncertainties, we carry out a Fisher matrix forecast for cosmological constraints from angular clustering in the redshift range $z = 0.2-0.95$. We quantify the cosmological constraints using a Figure of Merit (FoM) that measures the combined constraints on $\Omega_m$ and $\sigma_8$ in the context of $\Lambda$CDM cosmology. We find that the trade-off between sample size and photo-z precision is sensitive to 1) whether cross-correlations between redshift bins are included or not, and 2) the ratio of the redshift bin width $\delta z$ and the photo-z precision $\sigma_z$. When cross-correlations are included and the redshift bin width is allowed to vary, the highest FoM is achieved when $\delta z \sim  \sigma_z$. We find that for the typical case of $5-10$ redshift bins, optimal results are reached when we use larger, less precise photo-z samples, provided that we include cross-correlations. For samples with higher $\sigma_{z}$, the overlap between redshift bins is larger, leading to higher cross-correlation amplitudes. This leads to the self-calibration of the photo-z parameters and therefore tighter cosmological constraints. These results can be used to help guide galaxy sample selection for clustering analysis in ongoing and future photometric surveys.  
\end{abstract}




\section{Introduction}
\label{sec: Intro}

The large-scale structure (LSS) of the Universe carries rich cosmological information (e.g., \citealt{Dodelson2013}). Combined with early Universe observations of the cosmic microwave background (CMB) and expansion measurements using supernovae, galaxy surveys have significantly contributed to the establishment of the highly successful standard cosmological model. In this model, the energy density content of the Universe is dominated by two main constituents: dark energy in the form of a cosmological constant ($\Lambda$) \citep{Frieman2008,Amendola2010}  and Cold Dark Matter (CDM) (e.g., \citealt{Bertone2005,Bertone2016}). Current and future cosmological surveys will try to test the $\Lambda$CDM paradigm to high accuracy and search for new physics using a number of different probes.

One of the most important ways the information contained in the LSS can be extracted is by studying the statistical properties of the distribution of galaxies, which are (biased) tracers of the underlying matter field. The most widely used statistic is the 2-point galaxy correlation function in real space or, equivalently, its Fourier-space analogue, the power spectrum (e.g., \citealt{Baugh2000}).

Measuring the clustering of galaxies in three dimensional space, maintaining full radial information, requires the precise knowledge of their redshifts. While spectroscopic surveys can provide such accurate redshifts,  obtaining spectra is expensive. Thus, spectroscopic redshift is only measured for a small number of galaxies. This results in a noisy reconstruction of the galaxy field which in turn leads to a degradation of its statistical power. 

Photometric imaging surveys, on the other hand, provide a less accurate estimation of the redshifts of galaxies (photometric redshifts or photo-zs) from the color information obtained through multi-band photometry using a small number of filters (e.g., \citealt{Salvato2019}). This method allows one to estimate the redshifts of a significantly larger number of galaxies (at least an order of magnitude more), at the expense of losing most of the radial information \citep{Benitez2009,Asorey2012,Chaves2018}. For that reason, in photometric surveys, instead of the three dimensional galaxy distribution, one usually considers its projection in a number of redshift bins and measures the angular 2-point correlation function or the angular power spectrum (e.g., \citealt{Crocce2011, Asorey2016, Budav, Elvin_Poole2018}).  

One important aspect that affects the constraints on cosmology from clustering measurements in photometric surveys is the particular galaxy sample used: Briefly, the number of galaxies in a particular sample determines the uncertainty in the measurement of the angular power spectrum, with sparser samples leading to higher shot (or Poisson) noise. The magnitude of the photo-z parameters and their errors both affect the cosmological constraints; as we will discuss in detail, less precise photo-zs lower the amplitude of the power spectrum, while uncertainties in the photo-z parameters translate into uncertainties on the estimated cosmological parameters.

A common choice is to use a sample of Luminous Red Galaxies (LRGs).  LRGs are characterized by a uniform spectral energy distribution (SED)  with a sharp break at 4000{\AA} (rest frame) \citep{Eisenstein2001}. This feature allows the selection of this sub-sample of galaxies from the general population, as well as the estimation of their redshift (as it moves through the photometric filters) with high accuracy \citep{Padmanabhan2005, Rozo2016}. The benefit of using such a sample is that it provides very accurate and well-characterized photo-zs (probability distributions very close to Gaussian, with scatter of the order of $\sigma_{z}/(1 + z) \sim 0.01-0.02$). Furthermore, these galaxies are strongly clustered (high galaxy bias), resulting in a higher signal-to-noise ratio. However, LRGs constitute only a small sub-sample of the total number of galaxies available in a photometric survey. Thus, similar to the case of spectroscopic surveys, a relatively high noise term arises that affects the amount of cosmological information that can be recovered.

Another choice is to select all galaxies up to a limiting magnitude, where the survey provides a homogeneous coverage in depth. Flux-limited samples have been used e.g.,  in \cite{Crocce2016}. In this study, the sample is dominated by a population of blue galaxies, which are more populous, can be observed to much higher redshifts, but the accuracy in the determination of their photo-zs is much lower. These samples typically have a scatter of the order of $\sigma_z/(1+z) \sim 0.07 - 0.1$, while their probability distributions are not very-well characterized. Another choice is to select a sample that is dominated by, but is not limited to, red galaxies as a compromise between having small redshift errors and high number density of galaxies. Such a selection was performed in \cite{Crocce2019} , resulting in a sample that is appropriate for measurements of the Baryon Acoustic Oscillations (BAO) feature at high redshift.

Galaxy surveys have just started giving cosmological constraints comparable in precision to those obtained from CMB measurements \citep{DES1,Hildebrandt2018}. To enable the full exploitation of future clustering analyses and extract the maximum cosmological information, the galaxy sample selection process has to be optimized.  

Such an attempt of optimized sample selection is currently under investigation for the third year (Y3) analysis of the Dark Energy Survey (DES)\footnote{\url{https://www.darkenergysurvey.org/}} \citep{DES2005}, a wide-field photometric survey. The main focus is to expand the sample of LRGs used in the first year (Y1) analysis of the DES data. Briefly, the procedure consists of applying different flux limits in pre-defined redshift bins, obtain samples of different size, photometric redshift errors and biases and perform a Fisher forecast for each case in an attempt to  locate the sample that gives the best constraints \citep[in prep.]{Porredon2019}.
 
 The above approach is tailored to the Y3 DES data, specifically for the combination of galaxy clustering and galaxy-galaxy lensing and takes into account potential systematic effects and their uncertainties in a realistic way. This means that the results can be immediately used for Y3 data when an optimal sample is located. Our work also uses the characteristics of DES Y3 data as a baseline case, but we take a complementary approach by exploring a wider, more generic parameter space adopting a number of simple assumptions, focusing our analysis on galaxy clustering only. 

In this work, we assume that all samples follow a common underlying redshift distribution, that they share a common galaxy bias factor, and that the photo-z uncertainties are Gaussian. 
With these assumptions, we are able to explore a wide range of sample choices (with different sizes and photo-z errors) and evaluate their resulting cosmological constraints under one constant framework. In addition, the framework allows us to explore several different scenarios and aspects for a DES Y3-like clustering analysis: (1) only using the auto-correlation spectra (as in the DES Y1 analysis, \citealt{DES1}), (2) including the cross-spectra, (3) the effect of priors on the photo-z parameters, (4) the effect of the redshift bin width.

Throughout the paper,  we use the Fisher formalism to forecast the joint constraints on the set of parameters $\Omega_m - \sigma_8$ (fixing the galaxy bias parameter, $b_{\mbox{\scriptsize{g}}}$, to its fiducial value),  since these parameters can be best constrained from a photometric survey \citep{DES1}. We design the setup of this paper to approximately match the Y3 DES data characteristics; however our approach is  general and thus our conclusions can be used to guide the sample selection for clustering measurements in other surveys as well. 

The paper is organized as follows:  in Section \ref{sec: formalism} we review the formalism of  angular power spectra and cosmological forecasts we are going to use. In Section \ref{sec: survey_samples} we show examples of samples from DES data and how to select them. In Section \ref{sec: Samples_Constraints} we show how different samples give different cosmological constraints and we develop a simple model to describe a wide range of samples. We use that model in Section \ref{sec: Baseline} to forecast cosmological constraints using only auto-correlation spectra, and in Section \ref{sec: Cross_correlations} using both auto- and cross-correlation spectra. Finally, in Section \ref{sec: Bin_size} we study the dependence of our results on the redshift bin size. We summarize in Section \ref{sec: Summary}.

The fiducial cosmology adopted in this paper is flat $\Lambda$CDM with parameters (from DES Y1 + Planck + JLA + BAO \citep[TABLE II]{DES1}) $\Omega_m = 0.301, \,\,\sigma_8 = 0.798, \,\, \Omega_b = 0.0480, \,\, h = 0.682, \,\, n_s = 0.973$. Furthermore we assume sum of neutrino masses $\sum m_{\nu} = 0.06$ eV and reionization optical depth $\tau = 0.06$.

\section{Formalism}
\label{sec: formalism} 

\subsection{Angular Power Spectra}
\label{subsec: APS}

As we described in the introduction, photometric surveys measure, instead of the full three dimensional overdensity  field, the  two-dimensional projections of it in a series of tomographic redshift bins. The projected galaxy overdensity in a redshift bin $i$, $\delta^i_{\mbox{\scriptsize{gal}}}(\mathbf{\hat{n}})$, at an angular position $\mathbf{\hat{n}}$, can be written as:

\begin{equation}
\delta^i_{\mbox{\scriptsize{gal}}}(\mathbf{\hat{n}})  = \int_0^\infty  dz\, W^i(z) \delta_m(\chi(z)\mathbf{\hat{n}},z),
\end{equation}
where  $\delta_m(\chi(z)\mathbf{\hat{n}},z)$  is the matter overdensity at the three-dimensional position $\mathbf{x} = \chi (z)\mathbf{\hat{n}}$.  $\chi(z)$ is the comoving radial distance at redshift $z$ and $W^{i}(z)$ is the weighting kernel for galaxy clustering, which is given by:
\begin{equation}
W^{i}(z) = b(z)\frac{dN_g^i}{dz} \equiv b(z)\,n^i_g(z),
\end{equation}
where $dN_g^i/dz$ is the normalized redshift distribution of galaxies in that bin. The galaxy bias factor, $b(z)$, accounts for the fact that the observed galaxy overdensity field is a biased tracer of the underlying matter overdensity field.  We have assumed a linear and scale independent bias factor, i.e. that the two fields are related as $\delta_{\mbox{\scriptsize{gal}}} (z) = b(z) \delta_m(z)$ \citep{Fry1993}.

The projected overdensity in the $i$-th bin can be decomposed in spherical harmonics, $Y_{\ell m}$:
\begin{equation}
\delta^i_{\mbox{\scriptsize{gal}}}(\mathbf{\hat{n}}) = \sum_{\ell = 0}^{\infty}\sum_{m=-\ell}^{\ell} a_{\ell m}^i Y_{\ell m} (\mathbf{\hat{n}}).
\end{equation}
The angular cross power spectra between two bins $i$ and $j$, can be defined in terms of the harmonic expansion coefficients $a_{\ell m}$: 
\begin{equation}
\langle (a_{\ell m}^i)(a_{\ell' m'}^j)^* \rangle \equiv \delta_{\ell \ell'}\delta_{m m'} C_\ell^{ij}. 
\end{equation}
Using the Limber \citep{Limber1953,Loverde2008} and flat-sky approximations, we can write these  angular power spectra as: 
\begin{equation} 
\label{eq: APS}
C_\ell^{ij} = \int_0^\infty dz \frac{H(z)}{c}\frac{W^i(z)W^j(z)}{\chi(z)^2}P_{NL}\left(k=\frac{\ell+1/2}{\chi(z)}, z \right).
\end{equation}
Here, $H(z)$ is the Hubble parameter at redshift $z$ and $P_{NL}(k,z)$ is the non-linear, three-dimensional matter power spectrum at wavenumber $k$ and redshift $z$. Limber approximation is a good approximation for scales $\ell \goa 10$. In the above expression we have not included the effects of redshift space distortions. These can be shown to be negligible in the angular scales under consideration \citep{Padmanabhan2007}. We calculate the linear power spectrum using the CAMB module  \citep{Lewis2000} and then we use \texttt{Halofit} \citep{Takahashi2012} to get the nonlinear power spectrum.

\subsection{Forecasting Formalism}
\label{subsec: forecasts}

For our forecasts we rely on the standard Fisher recipe \citep[see e.g.,][]{Tegmark1997}. The Fisher matrix provides an approximation for the covariance matrix of the parameters a future experiment will try to measure.  Its elements, $F_{\mu \nu}$ are defined as the expectation value of the curvature of the log-likelihood with respect to the parameters of interest:
\begin{equation}
F_{\mu \nu} \equiv - \left\langle\left. \frac{\partial^2 \log {\cal{L}}}{\partial \theta_\mu \partial \theta_\nu} \right|_{\mathbf{\theta}=\mathbf{\theta_0}}\right\rangle,
\end{equation}
where $\theta_{\mu,\nu}$ the parameters we want to constrain,  and $\mathbf{\theta_0}$ their fiducial values. Then, a lower bound for the error on the measurement of the parameter $\theta_\mu$ can be estimated as: $\sigma_\mu \equiv \sigma(\theta_\mu) \geq \sqrt{(\mathbf{F}^{-1})_{\mu \mu}}$ (Cram\'er-Rao bound).

In the case where our data vector consists of the observed angular power spectra (including the cross-correlations) it can be shown that it takes the form (e.g., \citealt{Hu2004}):
\begin{equation}
\label{Eq: Fisher_mat}
F_{\mu \nu} = \frac{f_{\mbox{\scriptsize{sky}}}}{2} \sum_{\ell} (2\ell + 1)\mbox{Tr}\left[\mathbf{\hat{C}}_\ell^{-1} \frac{\partial \mathbf{\hat{C}}_{\ell}}{\partial \theta_\mu} \mathbf{\hat{C}}_\ell^{-1} \frac{\partial \mathbf{\hat{C}}_{\ell}}{\partial \theta_\nu} \right],
\end{equation}
where $f_{\mbox{\scriptsize{sky}}}$ is the fraction of the sky the survey covers and  $\mathbf{\hat{C}}$ is a matrix with elements the observed (i.e., including the shot noise) power spectra $\hat{C}_\ell^{ij}$:
\begin{equation}
\hat{C}_\ell^{ij} = C_\ell^{ij} + \delta^{ij}\frac{1}{\bar{n}_g^i},
\end{equation}
where $\bar{n}_g^i$ is the angular number density (number of galaxies per steradian) in the $i$-th redshift bin.

In the above formula we have taken into account all the cross-correlations between redshift bins. In the case where we consider only the auto-correlations, the formula reduces to:
\begin{equation}
F_{\mu \nu} = \sum_i \sum_{\ell_{\mbox{\scriptsize{{min}}}}}^{\ell_{\mbox{\scriptsize{{max,i}}}}} \frac{1}{\sigma^2_{\ell,i}}\frac{\partial C_\ell^{i}}{\partial \theta_\mu} \frac{\partial C_\ell^{i}}{\partial \theta_\nu},
\end{equation}
where:
\begin{equation}
\label{eq: delta_C}
\sigma_{\ell,i} \equiv 
\delta C_{\ell,i} = \sqrt{\frac{2}{f_{\mbox{\scriptsize{sky}}}(2\ell + 1)}}\left(C_\ell^{i} + \frac{1}{\bar{n}_g^i} \right).
\end{equation}
In the above $C_\ell^i \equiv C_\ell^{ii}$.  The first term in Eq. \eqref{eq: delta_C} is usually referred to as the cosmic variance term and the second as the shot noise term. The outer sum is over the redshift bins. We consider the same minimum multipole for each bin, $\ell_{\mbox{\scriptsize{min}}} = 10$, such that the Limber approximation holds \citep{Loverde2008}. The maximum $\ell$, corresponding to the minimum angular scale we consider for galaxy clustering measurements, is given by $\ell_{\mbox{\scriptsize{max,i}}} = k_{\mbox{\scriptsize{max}}}\chi(\bar{z})$, where $\bar{z}$ is the mean redshift of the bin. A common choice for the maximum (comoving) wavenumber, $k_{\mbox{\scriptsize{max}}}$ is to be $\sim 0.2 h$ Mpc$^{-1}$. This choice corresponds to a quasi-linear cutoff scale and thus does not require the modeling of non-linear scales. However, in DES the minimum (comoving) scale considered for clustering measurements is $R_{\mbox{\scriptsize{clustering}}} = 8$ Mpc $h^{-1}$ \citep{Krause2017}, which translates into a maximum $k_{\mbox{\scriptsize{max}}} = 2\pi/R_{\mbox{\scriptsize{clustering}}} \simeq 0.79 $ Mpc$^{-1}$ $h$. Here we will adopt a slightly more conservative choice,  $k_{\mbox{\scriptsize{max}}} = 0.6$ Mpc$^{-1}$ $h$, corresponding to $R_{\mbox{\scriptsize{clustering}}} \simeq 10$ Mpc $h^{-1}$, as in \cite{Krause2017a}. 

We can easily include Gaussian priors on the Fisher matrix. We simply add a prior matrix:
\begin{equation}
\label{eq: prior_F}
F_{\mu \nu} \to F_{\mu \nu} + F^P_{\mu \nu}.
\end{equation}
The prior matrix is diagonal with elements:
\begin{equation}
F^P_{\mu \nu} = \delta_{\mu \nu}\frac{1}{(\sigma^P_{\mu})^2},
\end{equation}
where $\sigma^P_{\mu}$ the prior error on the parameter $\theta_\mu$.

A commonly used metric to measure the constraining power of a given survey and dataset is the figure of merit (FoM) see e.g., \cite{Albrecht2006} . For a subset (of the total number) of cosmological parameters $\mathbf{\theta}$ we want to constrain, this is defined as:
\begin{equation}
\label{eq: FoM_1}
\mbox{FoM}_\mathbf{\theta} \equiv \frac{1}{ \sqrt{\det\left[ (\mathbf{F}^{-1})_\mathbf{\theta}\right]}},
\end{equation}
where the operation $(\mathbf{F}^{-1})_\mathbf{\theta}$ means that we first invert the total Fisher matrix $\mathbf{F}$ and then we keep only the rows and columns that refer to the parameters $\theta$.
In the case where this subset of cosmological parameters coincides with the set of cosmological parameters we leave free to vary, the above reduces to:
\begin{equation}
\mbox{FoM} = \sqrt{\det{\mathbf{F}}}.
\end{equation}
An intuitive way to understand the figure of merit, is to consider the marginalized posterior of the joint constraints of two parameters. Then, the figure of merit is inversely proportional to the area of the confidence ellipse of the constraints on these two parameters.

\subsection{Photometric Redshift Uncertainties }
\label{subsec: Photo-zs}

Following \cite{Ma2006}, we adopt a  simple Gaussian model for the photometric redshift uncertainties, characterized by a common scatter parameter, $\sigma_z$, that scales with redshift  as:
\begin{equation}
\sigma_z = \sigma_{z,0}(1+z),
\end{equation} 
and one constant photo-z bias parameter, $z^i_{\mbox{\scriptsize{b}}}$, per bin:
\begin{equation}
p^i(z_{\mbox{\scriptsize{ph}}}|z) = \frac{1}{\sqrt{2\pi} \sigma_z}\exp\left[-\frac{(z_{\mbox{\scriptsize{ph}}}-z-z^i_{\mbox{\scriptsize{b}}})^2}{2\sigma_z^2}\right].
\end{equation}

If the overall, normalized, redshift distribution of a sample is $dN_g/dz$, the galaxy clustering weighting kernel in a redshift bin $i$ can be written as:
\begin{equation}
\label{eq: W_i}
W^i(z) =b(z) \frac{\frac{dN_g}{dz}F^i(z)}{\int_0^\infty \frac{dN_g}{dz'}F^i(z') \,dz'} \, \equiv b(z) \frac{dN_g^i}{dz},
\end{equation}
where $dN_g^i/dz = \frac{\frac{dN_g}{dz}F^i(z)}{\int_0^\infty \frac{dN_g}{dz'}F^i(z') \,dz'} $ is the redshift distribution in the $i$-th bin.
$F^i(z)$ is a window function that gives the probability to include a galaxy in that bin.
For a spectroscopic survey, $F^i(z)$ is a top-hat function with limits those of each bin. For Gaussian photometric uncertainties, the window function becomes:
\begin{eqnarray}
\label{eq: F_i}
F^i(z)&=&\int_{z^i_{\mbox{\scriptsize{min}}}}^{z^i_{\mbox{\scriptsize{max}}}} dz_{\mbox{\scriptsize{ph}}} p^i(z_{\mbox{\scriptsize{ph}}}|z)  \\
\nonumber \\ 
&=&\frac{1}{2}\left[\mbox{erf}\left(x^i_{\mbox{\scriptsize{min}}}\right) - \mbox{erf}\left(x^i_{\mbox{\scriptsize{max}}}\right)\right] ,
\end{eqnarray}
with:
\begin{equation}
\label{eq: chi_def}
x^i_{\mbox{\scriptsize{min/max}}} \equiv \left(z - z^i_{\mbox{\scriptsize{min/max}}} -  z^i_{\mbox{\scriptsize{b}}} \right)/\sqrt{2}\sigma_z
\end{equation}
and
$z^i_{\mbox{\scriptsize{min/max}}}$ are the limits of the  $i$-th bin.

If we consider $m$ redshift bins, which  contain $N_{\mbox{\scriptsize{g}}}$ galaxies in total, the number of galaxies, $N_{\mbox{\scriptsize{g}}}^i$, in the $i$-th redshift bin is given by:

\begin{equation}
\label{eq: gal_in_bin}
N_{\mbox{\scriptsize{g}}}^i = N_{\mbox{\scriptsize{g}}} \frac{\int_0^\infty \frac{dN_{\mbox{\scriptsize{g}}} }{dz}F^i(z) \,dz}{{\sum_{j=1}^m} \int_0^\infty \frac{dN_{\mbox{\scriptsize{g}}} }{dz}F^j(z) \,dz}.
\end{equation}

\section{Survey and Samples}
\label{sec: survey_samples}

In this section we describe the Dark Energy Survey, the sample selection procedure, how we estimate photo-z uncertainties, redshift distributions and the assumptions we make about the nuisance parameters.

\subsection{The Dark Energy Survey}
\label{subsec: DES}

We consider the Dark Energy Survey (DES) as a typical example of a photometric galaxy survey. DES is a  survey that covered $ 5000$  deg$^2$ of the southern sky in five photometric filters, $grizY$, to a depth of $i \sim 24$ over a six year observational program using the 570-megapixel Dark Energy Camera (DECam) on the 4m Blanco Telescope at the Cerro Tololo Inter-American Observatory (CTIO) in Chile \citep{Flauger2015}.

For reasonable photo-z uncertainties, redshift distributions and number density of samples, we extract the relevant values from the first year (Y1) of DES observations, which are publicly available\footnote{\url{https://www.darkenergysurvey.org/the-des-project/data-access/}}. Our baseline analysis is for the third year (Y3) of DES, which we simulate by scaling the Y1 sky coverage ($\sim 1500$ deg$^2$) to Y3 ($ \sim 5000$ deg$^2$) while keeping everything else the same. Note that in practice the Y3 is expected to be deeper than Y1, so with our choice we underestimate the angular number density of the flux-limited sample (see Section \ref{subsubsec: Flux_lim}).  

The data set used has been processed and calibrated by the DES Data Managment system (DESDM) \citep{Sevilla2011} and finally curated, optimized and complemented in the Gold catalog (`Y1GOLD', \citealt{Drlica2018}). ``Bad" regions information is propagated to the 'object' level by using the \texttt{flags{$\_$}badregion} column in the catalog. Finally, individual objects identified as problematic are flagged as bad by using the \texttt{flags{$\_$}gold} column. To avoid imaging artifacts and bad regions we perform the cuts: \texttt{flags{$\_$}badregion < 4} and \texttt{flags{$\_$}gold = 0}.

Throughout this paper, when referring to magnitudes, we use \texttt{SExtractor}'s \citep{Bertin1996}  \texttt{MAG{$\_$}AUTO} quantities. We use photometric redshifts obtained with a Multi-Object Fitting (MOF) photometry \citep[section 6.3]{Drlica2018} and the Bayesian Photometric Redshifts (BPZ) algorithm \citep{Benitez}, a template fitting method that assigns a photo-z probability distribution function to each galaxy.

\subsection{Examples of sample selection}
\label{subsec: Sample_Selection}

We focus on three galaxy samples: a sample of LRGs (red-sequence
Matched-filter Galaxy Catalog or redMaGiC), a flux-limited sample and a sample for which, in addition to magnitude cuts, color cuts are applied in order to create a red-galaxies dominated sample.

\subsubsection{redMaGiC sample}
\label{subsubsec: redmag}

The galaxy sample used for clustering measurements in the first year (Y1) analyses of DES was a sub-sample of the DESY1 catalog, selected using  the redMaGiC algorithm \citep{Rozo2016}. The redMaGiC algorithm selects LRGs such that the redshift uncertainties are minimal ($\sigma_z/(1+z) <0.02$).  It selects galaxies above some luminosity threshold based on how well they fit a red sequence template. This template is  generated by the training of the redMaPPer cluster finder \citep{Rykoff2014}. The algorithm  produces for each galaxy a mean redshift prediction $z_{\mbox{\scriptsize{RM}}}$  and an uncertainty which is assumed to be Gaussian, characterized by its spread $\sigma_z$.

RedMaGiC computes color cuts necessary to produce a luminosity-thresholded sample of constant co-moving density. Higher luminosity thresholds would lead to a lower density. The sample we consider here is the same as in the Y1 analysis (but in the redshift range $z = [0.2-0.95]$), which is a combination of three redMaGiC galaxy samples, each of them defined to be complete down to a given luminosity threshold $L_{\mbox{\scriptsize{min}}}$, referred to as the high-density, high-luminosity and higher-luminosity samples. The corresponding luminosity thresholds and comoving densities for these samples are, respectively, $L_{\mbox{\scriptsize{min}}} =0.5L_*$, $L_*$ and $1.5L_*$, and $\bar{n} =10^{-3}, \,4 \times 10^{-4}$, and $10^{-4}$ galaxies$/(h^{-1} \mbox{Mpc})^3$ \citep{Elvin_Poole2018}.

\subsubsection{Flux-Limited sample}
\label{subsubsec: Flux_lim}

We consider a flux-limited sample defined by the magnitude cuts  \citep{Crocce2019}:
\begin{eqnarray}
17.5 < i <  22.
\end{eqnarray}
The limiting magnitude $i=22$ gives a complete sample over the DES Y1 footprint; a deeper limit would lead to inclusion of more objects in the sample, but the area over which it would be homogeneous it would be much smaller \citep[Figure 2]{Crocce2019}. We also remove the most luminous objects by applying the bright end cut. As mentioned in Section \ref{subsec: DES}, we expect DES Y3  to be homogeneous over a higher depth than Y1, which means that  we will be able to use a limiting magnitude higher than the one used here. Thus, our choice is a conservative one and underestimates the expected size of the flux-limited sample in Y3 analysis.

We also perform the following color cuts \citep{Crocce2016}:
\begin{eqnarray}
-1  <  g - r  <  3\\
-1  <  r- i  < 2.5\\
- 1 < i - z < 2.
\end{eqnarray}
These cuts remove color outliers that are either unphysical or from special samples (Solar System objects, high redshift quasars).

Furthermore, to avoid contamination of the galaxy clustering signal we have to remove the stars from the sample. To achieve this we use the default star-galaxy classification scheme that is based on \texttt{SExtractor}'s $i-$band coadd magnitude \texttt{spread{$\_$}model{$\_$}i} and its associated error \texttt{spreaderr{$\_$}model{$\_$}i}. Following \cite{Crocce2019},  we make our selection based on the combination:
\begin{equation}
\mbox{\texttt{spread{$\_$}model{$\_$}i} +(5.0/3.0)\texttt{spreaderr{$\_$}model{$\_$}i} > 0.007}, 
\end{equation}
which produces a sample with purity $97\% - 98\%$. We do not perform any additional masking.

\subsubsection{Color cuts}
\label{subsubsec: Color_cuts}

If we want to produce a sample that it is dominated by red galaxies (which have better photo-zs) but with less stringent selection criteria compared to the redMaGiC sample, we can define an additional sample constructed by imposing color cuts to the flux-limited sample. This is the approach used in  \cite{Crocce2019}.

As in \cite{Crocce2019}, in addition to the magnitude and color cuts of the previous section, we perform the following cuts as well:
\begin{equation}
(i-z) +2.0(r-i) >1.7.
\end{equation}
This selects red galaxies at high redshifts ($ z \gtrsim 0.6$) and was used in  \cite{Crocce2019} to define a galaxy sample to be used for BAO measurements. As we will see, the produced galaxy sample lies between the redMaGiC and flux-limited samples in terms of size and redshift uncertainty.

\subsubsection{Redshift distributions}
\label{subsub: distributions}

\begin{figure}
\centering
\includegraphics[width=\linewidth]{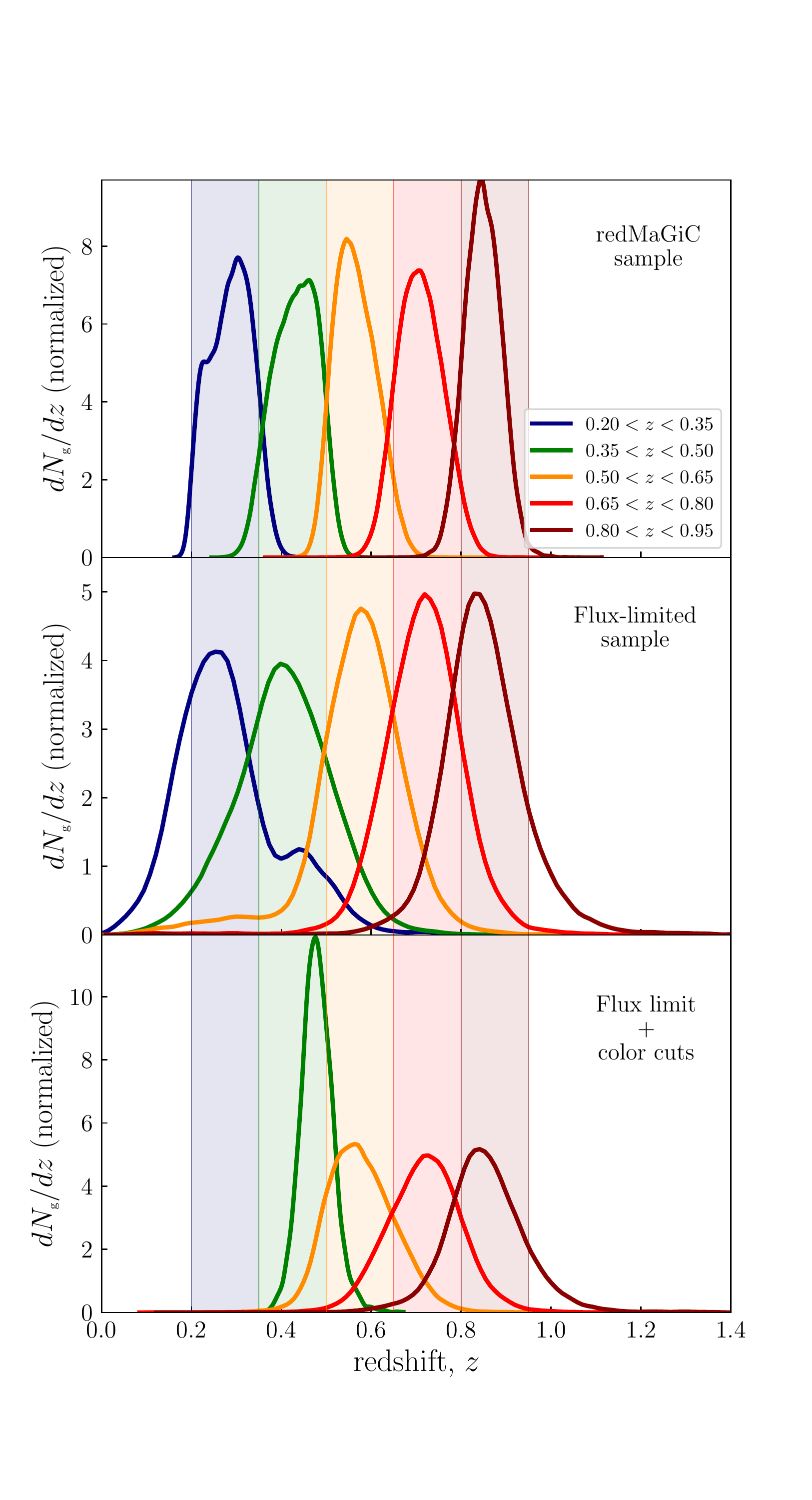} 
\caption{The normalized redshift distributions in five redshift bins between $z=0.2$ and $z=0.95$ of  the redMaGiC sample (top), a flux-limited sample (middle) and a sample which is defined through color cuts  on the flux-limited sample (bottom). The shaded regions correspond to the nominal limits of the redshift bins.}
\label{fig: Red_dist_data}
\end{figure}

\begin{table}
\centering
\caption{Number density and mean photo-z scatter of galaxies from DES Y1 data, in the five photometric bins between $z=0.20-0.95$ for the redMaGiC sample, the  flux-limited sample and a sample defined through color cuts.}
 \label{tab: bin_n_size}
 \begin{tabular}{lcc}
  \hhline{===}
   &   {\textbf{redMaGiC}}&  \\
  Area: 1321 deg$^2$ & & \\
  \hline
   bins&  $n_{\mbox{\scriptsize{gal}}}$ (deg$^{-2}$)   & ${\sigma}_z/(1+z)$ \\
  \hline
  0.20 -- 0.35 & 71.8  & 0.015 \\
  0.35 -- 0.50 & 148.0 & 0.017 \\
  0.50 -- 0.65 &  157.3 & 0.015 \\
  0.65 -- 0.80 &  85.7 & 0.020 \\
  0.80 -- 0.95 &  22.4 & 0.015 \\
  \hline
  \hhline{===}
   & {\textbf{Flux-limited}} &  \\
   Area: 1536 deg$^2$ & & \\
   \hline
   bins&  $n_{\mbox{\scriptsize{gal}}}$ (deg$^{-2}$)   & ${\sigma}_z/(1+z)$  \\
  \hline
  0.20 -- 0.35 & 2274.3 & 0.081  \\
  0.35 -- 0.50 & 5884.9  & 0.076  \\
  0.50 -- 0.65 & 2338.3 & 0.075 \\
  0.65 -- 0.80 & 2081.6 & 0.056 \\
  0.80 -- 0.95 & 931.4 & 0.057 \\
  \hline
  \hhline{===}
  & {\textbf{Color cuts}} &  \\
   Area: 1536 deg$^2$ & & \\
   \hline
   bins&   $n_{\mbox{\scriptsize{gal}}}$ (deg$^{-2}$)    & ${\sigma}_{z}/(1+z)$  \\
  \hline
  0.20 -- 0.35 & -- & -- \\
  0.35 -- 0.50 & 105.7  & 0.023 \\
  0.50 -- 0.65 & 1106.3 & 0.040 \\
  0.65 -- 0.80 & 1497.7 & 0.044 \\
  0.80 -- 0.95 & 769.8 & 0.044 \\
  \hline
 \end{tabular}
\end{table}

For the three choices of samples mentioned above we use galaxies in the redshift range $z=0.2$ to $z=0.95$. We bin them into five bins of width $\delta z  = 0.15$. This bin width was used in the  DES Y1 clustering analysis using the redMaGiC sample \citep{Elvin_Poole2018}. For the flux-limited and color cuts-defined samples, each galaxy is assigned to a bin according to its mean photo-z estimate \texttt{bpz$\_$zmean$\_$mof}. For the redMaGiC sample we use the mean estimate  \texttt{ZREDMAGIC}. The angular number density of galaxies per bin for the three samples is shown in the second column of Table \ref{tab: bin_n_size}.

To get the redshift distributions of the flux-limited and color cuts-defined  galaxy samples in each bin, we assign to each of them a randomly selected photometric redshift value from their photo-z pdf,  \texttt{bpz$\_$zmc$\_$mof}. For the redMaGiC sample we select a redshift drawn from a Gaussian with mean given by the \texttt{ZREDMAGIC} and width given by the error estimate,  \texttt{ZREDMAGIC$\_$E}. The resulting normalized redshift distributions are shown in Fig. \ref{fig: Red_dist_data}. 

Mathematically, this process is not strictly correct, since it returns an error-convolved version of the underlying distribution instead of the true distribution. However, at the level of the current surveys, this approximation is appropriate and has been used for most of the recent analyses (e.g., \citealt{Elvin_Poole2018,Hoyle2018}).

For each sample we estimate one redshift uncertainty scatter parameter, $\sigma_{z,0}$, per bin, by fitting a Gaussian to the histogram of their randomly selected photometric values (after subtracting the mean and dividing by $1+z$, where $z$ the estimated redshift). The values of the estimated mean $\sigma_z/(1+z)$, can be found in the third column of Table \ref{tab: bin_n_size}.

\subsubsection{Scaling to Y3 footprint}
\label{subsub: scaling}

The Y1 Gold catalog, after masking for bad regions, that we used to produce the flux-limited and color cuts-defined samples covers a footprint of $1536$ deg$^2$, while the redMaGiC catalog covers $1321$ deg$^2$. Using these numbers and the number of galaxies per bin from the Y1 data we estimeted the angular number densities presented in Table \ref{tab: bin_n_size}.

From now on, we perform our calculations by keeping the number density of each sample fixed to the average Y1 value, but scaling up to the expected footprint of $5000$ deg$^2$ ($f_{\mbox{\scriptsize{sky}}} \cong 0.12$) for DES Y3. The number of galaxies and average photo-z scatter $\sigma_{z,0}$ (defined through $\sigma_{z,0} = \sigma_z/(1+z)$) in the $[0.2 - 0.95]$ redshift range, as resulting from the figures presented in Table \ref{tab: bin_n_size}, for the three samples are:

$\bullet$ redMaGiC: $N_{\mbox{\scriptsize{g}}} \cong 2.46 \times 10^6$, $\sigma_{z,0} = 0.017$.

$\bullet$ Flux-limited: $N_{\mbox{\scriptsize{g}}} \cong 6.75 \times 10^7$, $\sigma_{z,0} = 0.073$.

$\bullet$ Color cuts-defined: $N_{\mbox{\scriptsize{g}}} \cong 1.74 \times 10^7$, $\sigma_{z,0} = 0.042$.

From now on we use these numbers to characterize these three samples.

\section{From Samples to Constraints}
\label{sec: Samples_Constraints}

\subsection{Photo-z accuracy vs. sample size}
\label{subsec: Phot_red_acc}

\begin{figure*}
\centering
\includegraphics[width=\textwidth]{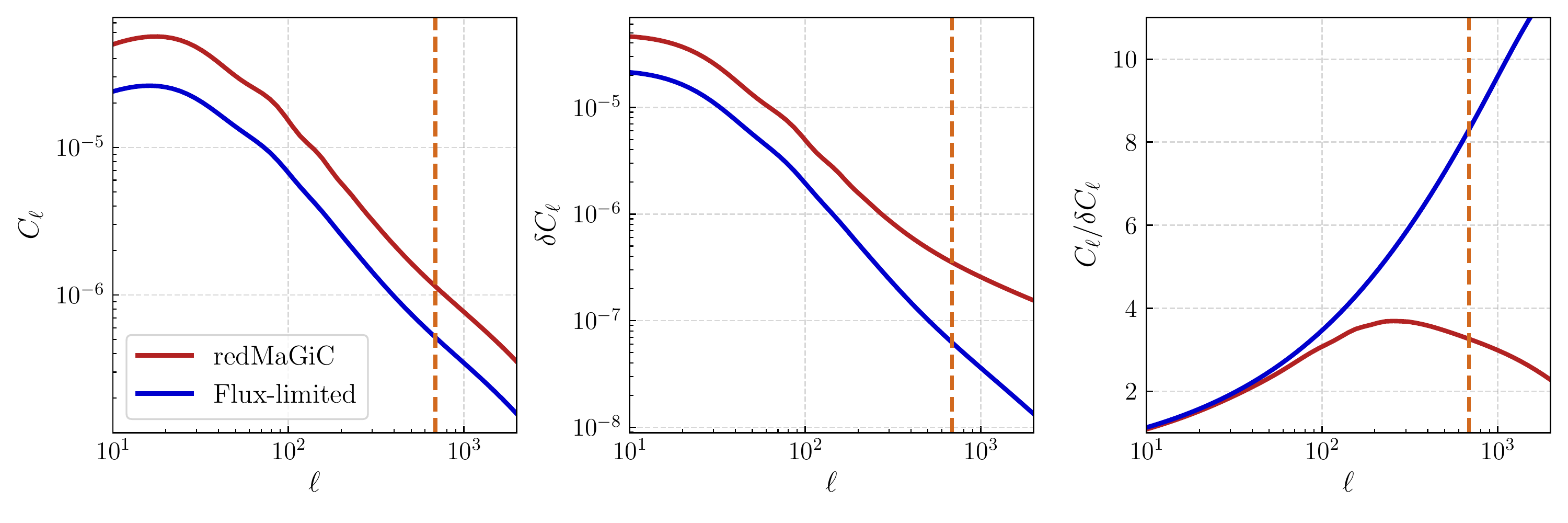} 
\caption{\textit{Left panel:} The angular power spectrum, $C_\ell$, of a redMaGic (red)  and a flux-limited  (blue) sample, as defined in Sec.  \ref{subsec: Phot_red_acc}, in a $0.35 < z <0.50$ redshift bin. \textit{Central panel:} The error on the angular power spectrum, $\delta C_\ell$ for the same samples. \textit{Right panel:} The ratio $C_\ell/\delta C_\ell$ for the two samples, presents the signal-to-noise we have in the two cases. In all three panels we also show (brown dashed vertical line) the maximum $\ell$ cutoff scale we use for our forecasts in that bin,  calculated to be $\ell = 687$ (see discussion in Section \ref{subsec: forecasts}). For both samples we have assumed the same linear galaxy bias $b_g = 1 + \bar{z}$, where $ \bar{z}$ is the mean redshift of the bin (here $\bar{z} = 0.425$). See also the discussion in Section  \ref{subsubsec: Gal bias}. }
\label{fig: APS_and_Error}
\end{figure*}

\begin{figure}
\centering
\includegraphics[width=0.75\columnwidth]{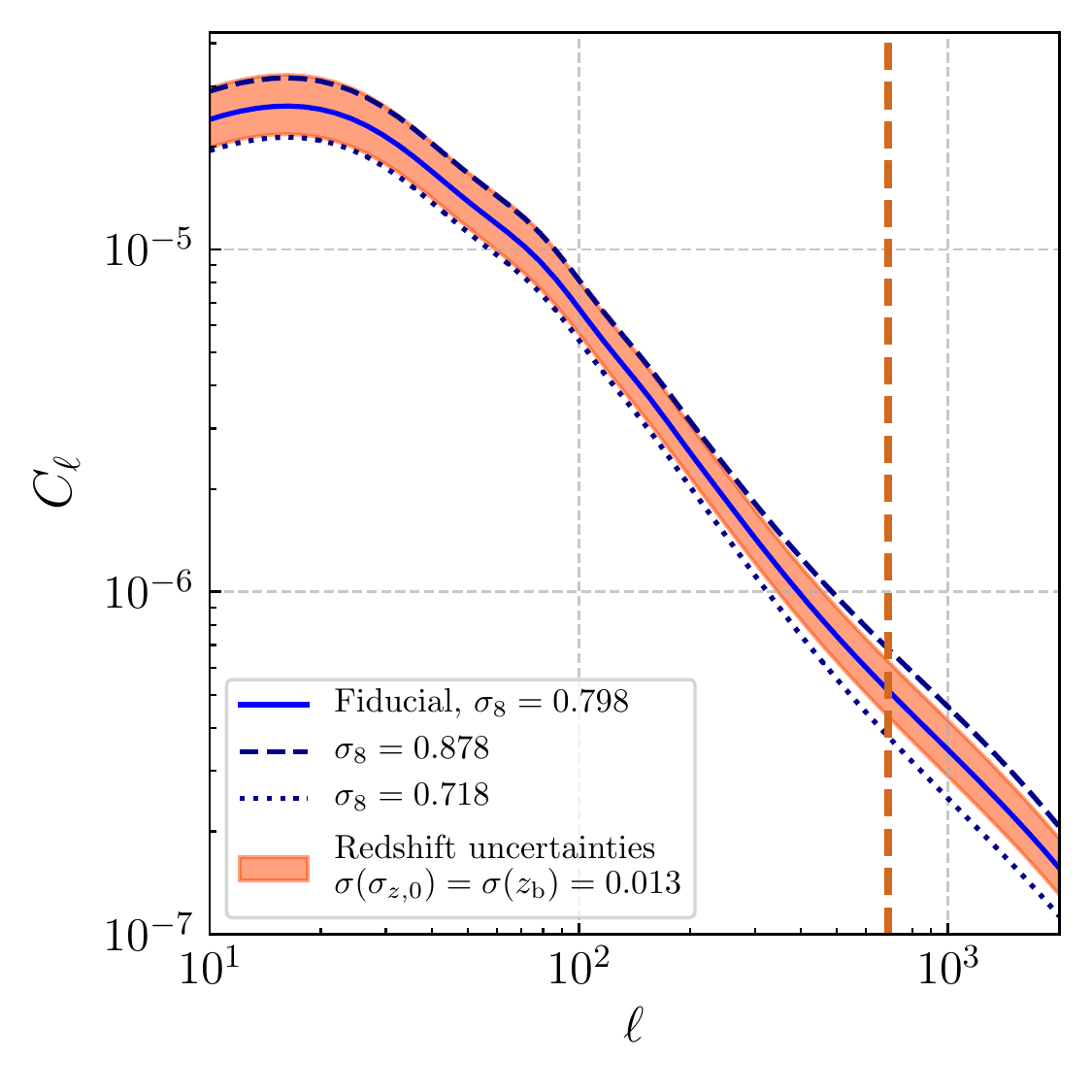} 
\caption{The angular power spectrum of the flux-limited sample in the $0.35-0.50$ redshift bin assuming $\sigma_8 = 0.798$ (solid blue line) and $\sim10\%$ higher and lower values for $\sigma_8$ (dashed and dotted blue lines, accordingly). The shaded red region corresponds to the allowed range for the power spectrum if we assume uncertainties on the photo-z parameters of the order of   $\sigma(\sigma_{z,0}) = \sigma(z_{\mbox{\scriptsize{b}}}) = 0.013$.}
\label{fig: red_uncert_eff}
\end{figure}

In this Section we discuss how the sample size, the photo-z accuracy and the level of calibration of the photo-z parameters compete and affect the information that can be extracted from a sample, and thus the cosmological constraints one can get by studying their clustering properties.

Let us consider a single redshift bin of width $\delta z = 0.15$ in the range $0.35 < z < 0.50$ (the second of the five redshift bins presented in the previous section) and two specific samples:\\
$\bullet$ One with  $N_{\mbox{\scriptsize{g}}}= 0.74 \times 10^6$, $\sigma_{z,0} = 0.017$. This choice corresponds to the  redMaGiC sample in that bin.\\
$\bullet$ One with $N_{\mbox{\scriptsize{g}}}= 2.94 \times 10^7$ and $\sigma_{z,0} = 0.076$. This choice corresponds to the flux-limited  sample.

For both samples we assume a fiducial value for the photometric redshift bias, $z_{\mbox{\scriptsize{b}}} = 0$. As we explained these two samples represent two limiting cases of possible samples selected in photometric surveys: the redMaGiC sample is a small sample with very accurate and secure photo-zs, while the flux-limited sample contains all galaxies up to a limiting magnitude and thus it has larger photo-z uncertainties. 

 In Fig. \ref{fig: APS_and_Error} we plot the predicted angular power spectrum, (defined in Eq. \eqref{eq: APS}, for $i=j$), its error $\delta C_\ell$ (defined in Eq. \eqref{eq: delta_C})(center) and their ratio (right) for the two samples defined above. 

The amplitude of the angular power spectrum, for fixed bin width and galaxy bias parameter, depends on the photometric redshift error (see Eqs. \eqref{eq: APS},\eqref{eq: W_i},\eqref{eq: F_i}); higher photometric error means a broader redshift distribution and thus lower amplitude. So, as expected, the redMaGiC sample that has better photo-zs gives an angular power power spectrum of higher amplitude than the flux-limited sample. 

The error on the power spectrum, $\delta C_{\ell}$, has two contributions: cosmic variance, because of the limited  number of observable modes in low $\ell$, and the shot (Poisson) noise $1/ \bar{n}_{\mbox{\scriptsize{g}}}$ that  has to do with the fact that we use a discrete number of points (galaxies) to measure the statistical properties of the underlying field. The redMaGiC sample has more precise photo-zs, but it is also sparser and thus the shot noise is higher. For low $\ell$, the main source of noise is the cosmic variance and the difference in the two samples has to do with the difference in the $C_\ell$. For higher $\ell$ the shot noise dominates and we see that  the spread in the $\delta C_{\ell}$ for the two samples increases, with the redMaGiC sample having higher error, as expected.  

The resulting signal to noise ratio, $C_\ell/\delta C_\ell$,  for the two samples can be seen in the right panel of Fig. \ref{fig: APS_and_Error}. For low $\ell$, as we have said, the dominant source of noise is cosmic variance, and so the signal-to-noise ratio is the same for the two samples. In higher $\ell$, the signal-to-noise ratio of the flux-limited sample is much higher than that of the  redMaGiC sample; the accuracy of the photo-zs is not enough to compensate for the high shot-noise of the sample. 

So far we have assumed a perfect knowledge of the photo-z uncertainties of the two samples. In practice though these uncertainties are known to a finite confidence level. We can characterize that by introducing uncertainties in the two photo-z parameters (scatter and bias) in the Gaussian model. 

The priors on the photo-z parameters can significantly impact the cosmological information of a photometric galaxy sample. In Fig. \ref{fig: red_uncert_eff} we plot the predicted angular power spectrum of the flux-limited sample in the $0.35 < z < 0.50$ redshift bin. The solid blue line assumes no redshift uncertainty and fiducial value for the power spectrum normalization parameter $\sigma_8 = 0.798$.  The dashed blue line corresponds to a cosmology with $ \sim 10\%$ higher $\sigma_8 , \,\, \sigma_8 = 0.878$ and the dotted blue line corresponds to a value of $\sigma_8$ that is $\sim 10\%$ lower than the fiducial, $\sigma_8 = 0.718$. The red shaded region shows the effect of including uncertainties on the photo-z parameters $\sigma(\sigma_{z,0}) = \sigma(z_{\mbox{\scriptsize{b}}}) = 0.013$. As we can see, the resulting uncertainty in the reconstruction of the angular power spectrum is comparable to a $\sim 10\%$ change of the parameter $\sigma_8$ across all scales. Thus an uncertainty in the photo-z parameters finally results  in an uncertainty on the value of the cosmological parameters we can measure from angular clustering (see also e.g., \citealt{Hearin2012}).

\subsection{Mapping data to model}

In the previous section we showed in detail how redshift distributions can be constructed from DES data, for three specific samples. To describe a broader range of samples, we need to develop a simple but more general model of the redshift distribution of samples with different photo-z uncertainty. That way we will be able to investigate, in a unified way, the trade-offs between photo-z uncertainties and sample size, the importance of cross-correlations between tomographic bins, and the effect of different bin widths.

\begin{figure}
\centering
\includegraphics[width=1.0\columnwidth]{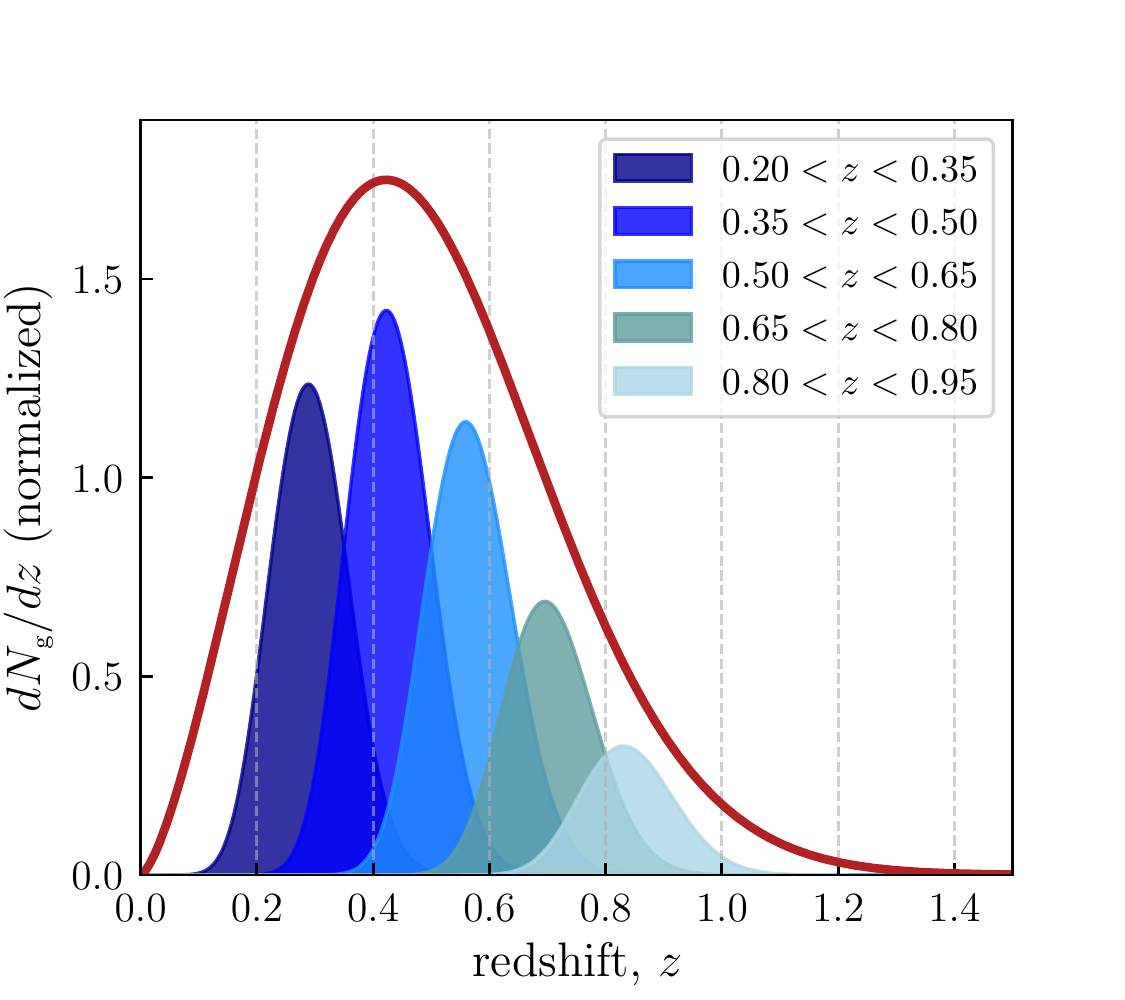} 
\caption{The overall, normalized, redshift distribution considered here (red line)  and the redshift distributions in the five bins (0.20-0.35, 0.35-0.50, 0.50-0.65, 0.65-0.80, 0.80-0.95), assuming a photometric redshift error $\sigma_{z,0}=0.04$.}
\label{fig: red_dist_mod}
\end{figure}

Here we present the assumptions behind the model we are going to use in the following sections. We also show how our model performs in forecasting the constraints from the three samples discussed in the previous section, compared to the forecasts we get when we use the redshift distributions and systematic biases obtained directly from the DES data.

\begin{figure*}
\centering
\subfigure[]{\includegraphics[width=0.7\columnwidth]{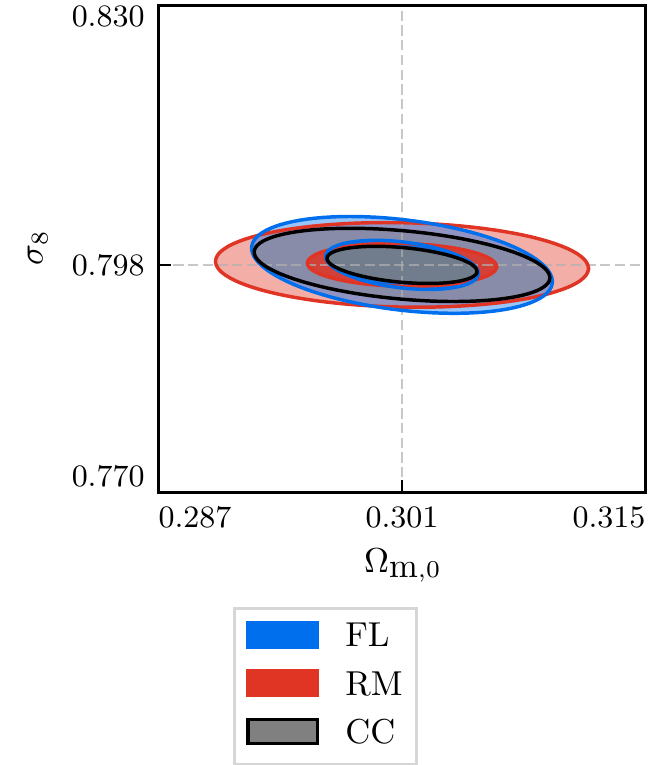}}
\subfigure[]{\includegraphics[width=0.7\columnwidth]{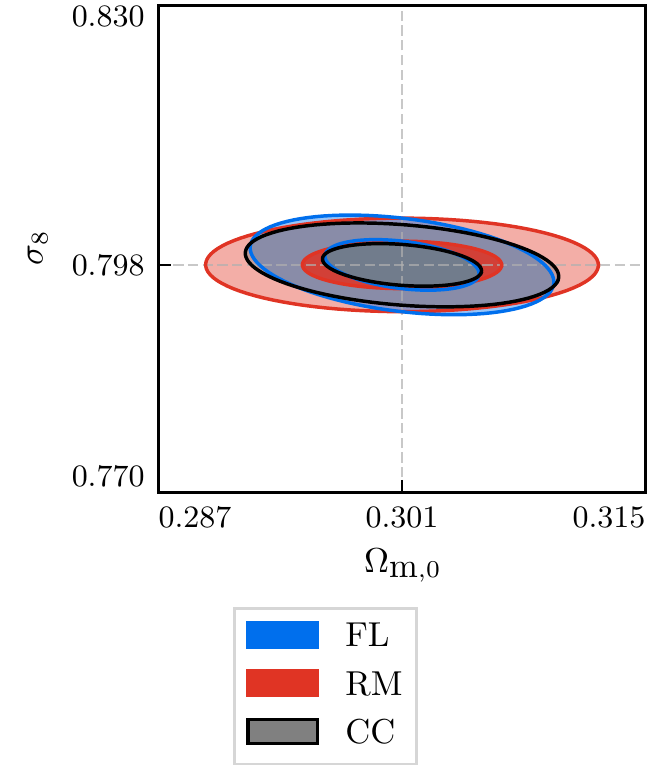}}
\subfigure[]{\includegraphics[width=0.7\columnwidth]{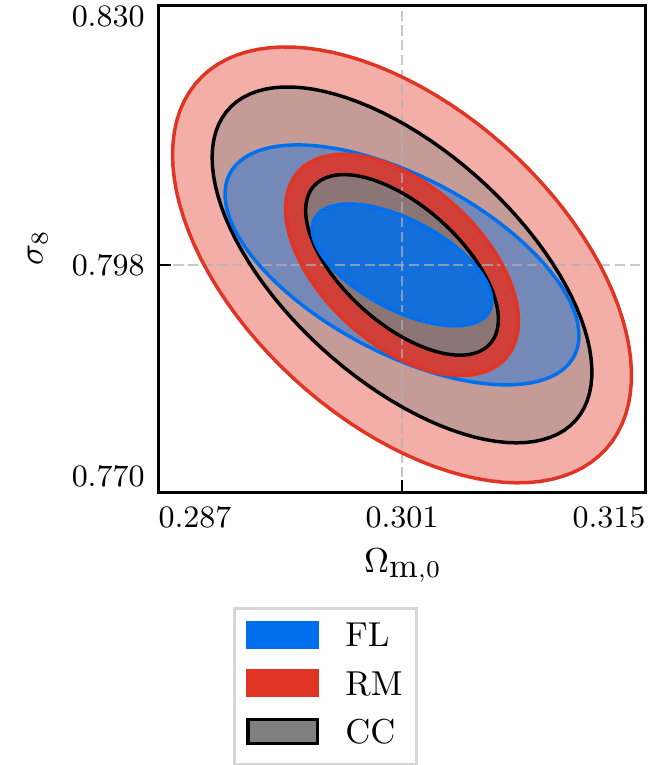}}
\caption{ Forecasts of constraints on the cosmological parameters $\Omega_{m} - \sigma_8$ using the redMaGiC (red),  flux-limited (blue) and color cuts-defined (gray) samples, described in Section \ref{subsec: Sample_Selection}. {\textit{Panel (a):}} Using the redshift distributions obtained from the DES Y1 data. {\textit{Panel (b):}} Using a Gaussian photo-z model with a common underlying redshift distribution for the three samples and keeping the photo-z scatter parameter fixed.  {\textit{Panel (c):}} As in panel (b), but having the photo-z scatter parameter free, with some priors (see text).}
\label{fig: Data_and_model}
\end{figure*}

\subsubsection{Underlying redshift distribution}

We adopt, for simplicity, a common underlying  redshift distribution for all samples, of the form \citep{Efstathiou1991, Smail1994}:
\begin{equation}
\label{eq: red_dist}
\frac{dN_{\mbox{\scriptsize{g}}}}{dz}(z) = \frac{\beta}{z_0^{1+a}\Gamma\left[ \frac{1+ \alpha}{\beta}\right]}z^\alpha \exp \left[-\left( \frac{z	}{z_{0}^{}}\right)^\beta \right],
\end{equation}
where $z_0, \,\alpha,\, \beta$ are parameters that determine the depth and the shape of the distribution, while the prefactor $ \frac{\beta}{z_0^{1+a}\Gamma\left[ \frac{1+ \alpha}{\beta}\right]}$ ensures that the distribution is normalized: $\int_0^{\infty} \frac{dN_{\mbox{\scriptsize{g}}}}{dz}dz = 1$. We obtain the values of the parameters by fitting Eq. \eqref{eq: red_dist} to the flux-limited sample from DES Y1 data; we have: $z_0 = 0.50, \, \alpha=1.47, \, \beta = 2.09$. In Fig. \ref{fig: red_dist_mod}  this overall redshift distribution is plotted with the thick red line.

\subsubsection{Redshift uncertainties}

As we described in Section \ref{subsec: Photo-zs}, we model redshift errors as  following a Gaussian distribution with a common scatter ($\sigma_{z,0}$) and one bias parameter ($z_{\mbox{\scriptsize{b}}}$) per bin. This approach is usually adopted in the relevant literature concerning cosmological forecasts from photometric surveys. In order to calculate the redshift distribution per redshift bin (which is the input for the calculation of the power spectra) we convolve the overall redshift distribution with the Gaussian PDF, Eqs. \eqref{eq: W_i}-\eqref{eq: F_i}. That way, uncertainties in the photo-z parameters are propagated into uncertainties of the redshift distributions.

In the DES Y1 analyses, the uncertainty in the photometric redshifts was taken into account by introducing one shift parameter $\Delta z^i$ per redshift bin, such that the galaxy redshift distribution in the $i$-th bin,  $dN^i_{\mbox{\scriptsize{g}}}/dz \equiv n_g^i(z) $, is written as \citep{Hoyle2018, Elvin_Poole2018}: 
\begin{equation}
n_g^i(z) = \hat{n}^i_g(z - \Delta z^i)
\label{eq: shifts}.
\end{equation}
$\hat{n}^i_g$ is the estimated redshift distribution, obtained from stacking the photo-z of  galaxies, in a manner similar to what we described in Section \ref{subsub: distributions}. This is equivalent (if we assume photo-zs approximately Gaussian) to fixing the photo-z scatter parameter to its fiducial value and leaving the photo-z biases free, and including some prior on them. 
For DES Y1, it was shown that the cosmology results were not sensitive at sub-sigma level to changes in the shape of $n_g^i(z)$ aside from an overall shift in redshift \citep{Hoyle2018, Elvin_Poole2018}; here, to allow more freedom, we leave both the photo-z scatter and biases free, with priors of the form $\sigma(\sigma_z) = \sigma(z_b^i) \propto \sigma_z$, i.e. proportional to the photo-z scatter (e.g., \citealt{Hearin2010}, see also \ref{subsec: Depend_on_priors}). We compare results from the two approaches in \ref{subsub: example} and in Appendix \ref{sec: shift_and_Gaussian}.

\subsubsection{Galaxy bias}
\label{subsubsec: Gal bias}

For our forecasts we have to assume a model for the galaxy bias, $b(z)$, that connects the matter to the galaxy overdensity field. 
Linear bias evolves with redshift (e.g.  \citealt{Fry1996, Clerkin2015}) and also depends on the specific sample being studied, with samples dominated by redder galaxies having higher bias than those dominated by bluer galaxies. 

Fitting a linear function of the form $b(z) = a + cz$ to the values of the galaxy bias found for the redMaGiC sample in five redshift bins used for clustering measurement in DES Y1 \citep{Elvin_Poole2018}, we find:
\begin{equation}
\label{eq: RM_bias}
b_{\mbox{\scriptsize{RM}}}(z) \cong 1.167 + 1.013z.
\end{equation}
Similarly, fitting a linear function to the measurements of the galaxy bias of the flux-limited sample from the  Science Verification (SV) data of DES \citep{Crocce2016}, which is dominated by blue galaxies, we find:
\begin{equation}
\label{eq: FL_bias}
b_{\mbox{\scriptsize{FL}}}(z) \cong 0.816 + 0.795z.
\end{equation}

Here, we adopt, for simplicity, a common  galaxy bias for all samples, that evolves with redshift as:
\begin{equation}
\label{eq: Mod_bias}
b(z) = 1 + z.
\end{equation}
The above is an approximation to Eqs. \eqref{eq: RM_bias},\eqref{eq: FL_bias} that is accurate enought for our purposes.
Furthermore, we assume a constant bias per bin, equal to $b_g^i = 1 + \bar{z}^i$, where $\bar{z}^i$ the mean redshift of the $i$-th redshift bin. Note that this approximation, of a constant bias per bin, without redshift evolution, may break down for samples with high $\sigma_{z,0}$; for such samples, due to the high redshift uncertainty there will be a mixing of populations from different (true) redshifts. 

Note that we fix the bias to make the resulting FoM easier to interpret. In reality, if we do not have any prior knowledge of the galaxy bias, galaxy clustering only constrains the combined parameter $\sigma_8 b_g$.

\subsubsection{A specific example}
\label{subsub: example}

Before using our simple model to forecast constraints in a range of different samples, we would like to test its performance over a more detailed description for the three samples presented in Section \ref{subsec: Sample_Selection}. 

In Fig. \ref{fig: Data_and_model} we present the forecast constraints on the cosmological parameters $\Omega_{m}$ and $\sigma_8$ from the redMaGiC, flux-limited and color cuts-defined samples, in five bins between $z=0.2$ and $z=0.95$, using auto-correlations only.

To get the results in panel (a) we use the redshift distributions of the three samples obtained from data and presented in Fig. \ref{fig: Red_dist_data}, and galaxy bias parameters according to Eqs. \eqref{eq: RM_bias} and \eqref{eq: FL_bias} for the redMaGiC and flux-limited samples. We introduce five nuisance shift parameters $\Delta z^i$ (as in Eq. \eqref{eq: shifts}), with priors $\sigma(\Delta z^i) = 0.007, 0.013, 0.009$ for the redMaGiC, flux-limited and color cuts-defined samples \citep{DES1}. 

In panel (b) we use the model we described earlier, adopting the same overall redshift distribution, Eq. \eqref{eq: red_dist}, for all three samples, and a common galaxy bias parameter according to the model of Eq. \eqref{eq: Mod_bias}. We use Gaussian photo-zs with scatter $\sigma_{z,0} = 0.017, 0.073, 0.042$, as described in Section \ref{subsub: scaling}, and one photo-z bias parameter  per bin, with value $z_{\mbox{\scriptsize{b}}}^i = 0$. To compare with the previous case in an equal footing, we fix the photo-z scatter to its fiducial value, while we impose the same priors as before to the photo-z bias parameters and marginalize over them. The differences in the forecast constraints on the two cosmological parameters between the  data-based and model-based cases are of the order of $\sim 3\% - 10\%$ for all samples, which is sufficient for the purpose of this work. For the remainder of the paper, we therefore adopt the assumptions that all samples have a common redshift distribution and Gaussian photo-zs.

In panel (c) we use the same model as in (b), but now we leave the photo-z scatter free, imposing on it the same priors as on the photo-z bias and marginalizing over them. Compared to the previous results, we note two main differences: First, now we get significantly worse constraints on $\sigma_8$ (for the redMaGiC sample $\sigma(\sigma_8)$ is $\sim 3$ times larger). This is expected since we add one more nuisance parameter that is highly degenerate with $\sigma_8$ (see the discussion in Section \ref{subsec: Phot_red_acc}). Furthermore we see that, while in the previous cases all samples seem to perform equally well on $\sigma_8$, here the flux-limited sample gives much better constraints than the redMaGiC sample, resulting in a much higher FoM (FoM$_{\mbox{\scriptsize{FL}}}$/FoM$_{\mbox{\scriptsize{RM}}} \sim 2.5$ here, while FoM$_{\mbox{\scriptsize{FL}}}$/FoM$_{\mbox{\scriptsize{RM}}} \sim 1.2$ in (b)). 

From now on we are going to use the model with free photo-z scatter parameter, as in panel (c), but we show how some of our results change when we fix it in Appendix \ref{sec: shift_and_Gaussian}. The cases in Appendix \ref{sec: shift_and_Gaussian} are interesting since current state-of-the-art analyses do not in general leave the photo-z scatter free. Our approach of leaving photo-z scatter free is a more general case of the analysis.

\begin{figure*}
\centering
\subfigure[]{\includegraphics[width=\columnwidth]{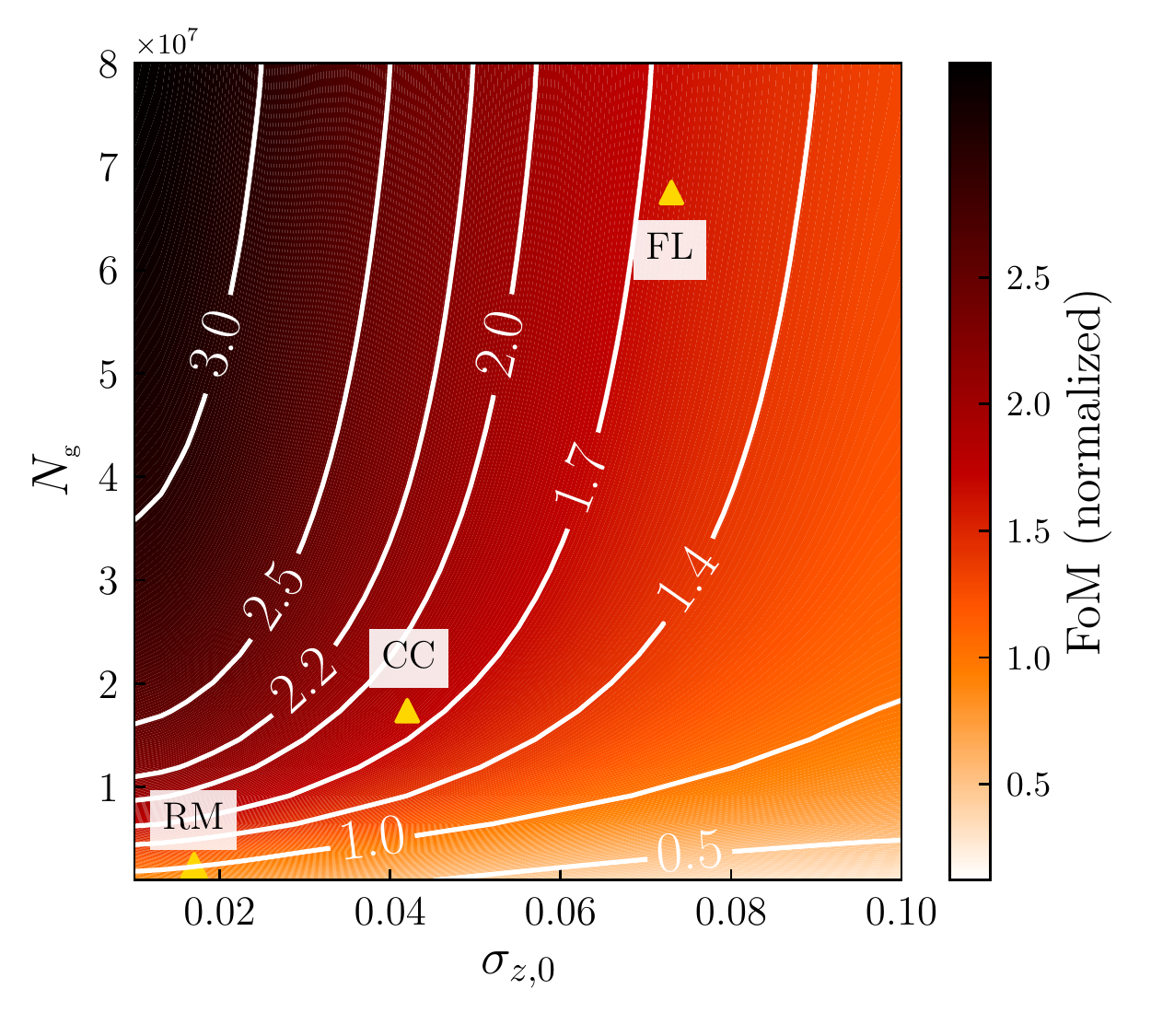}}
\subfigure[]{\includegraphics[width=\columnwidth]{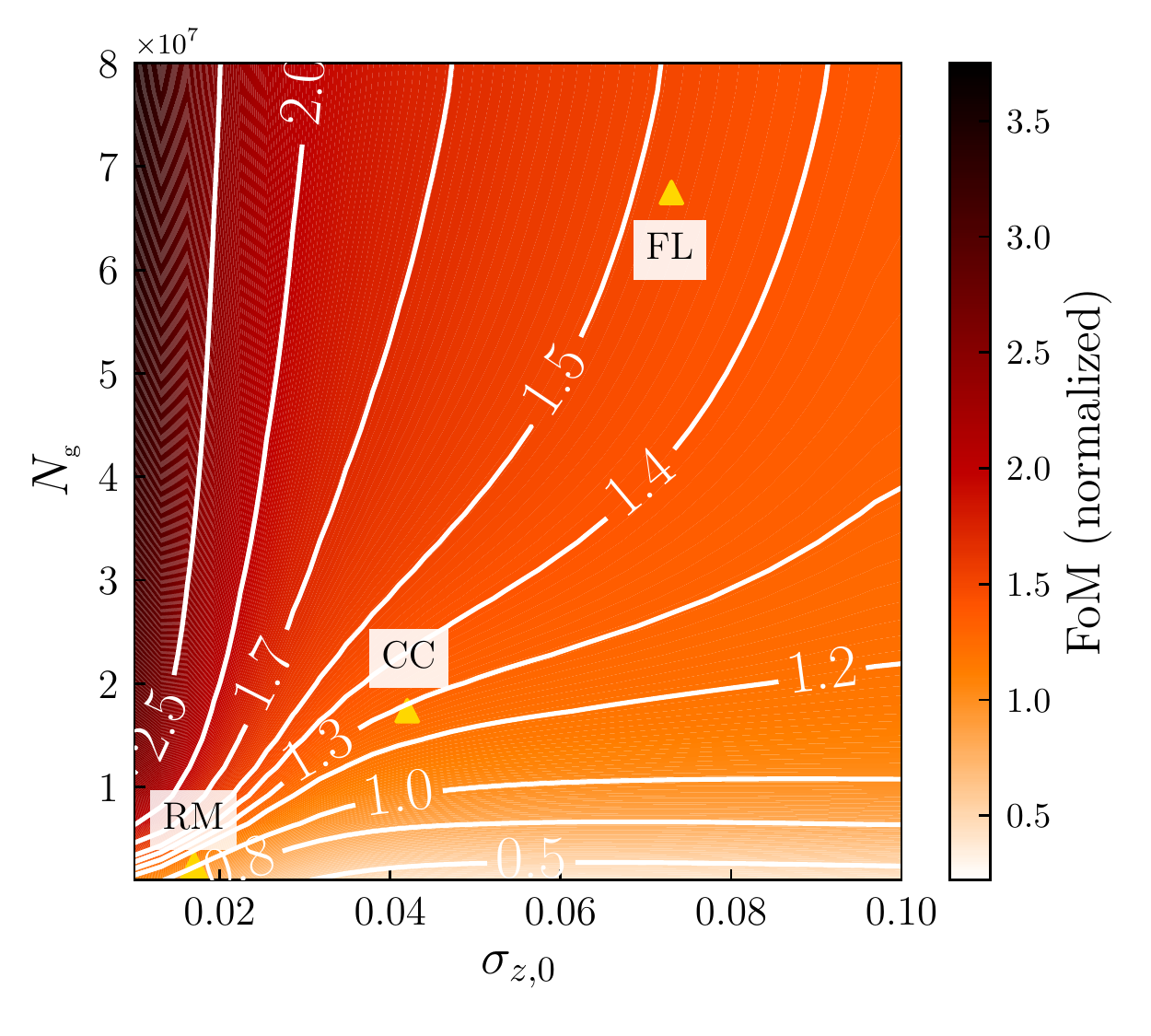}}

\subfigure[]{\includegraphics[width=\columnwidth]{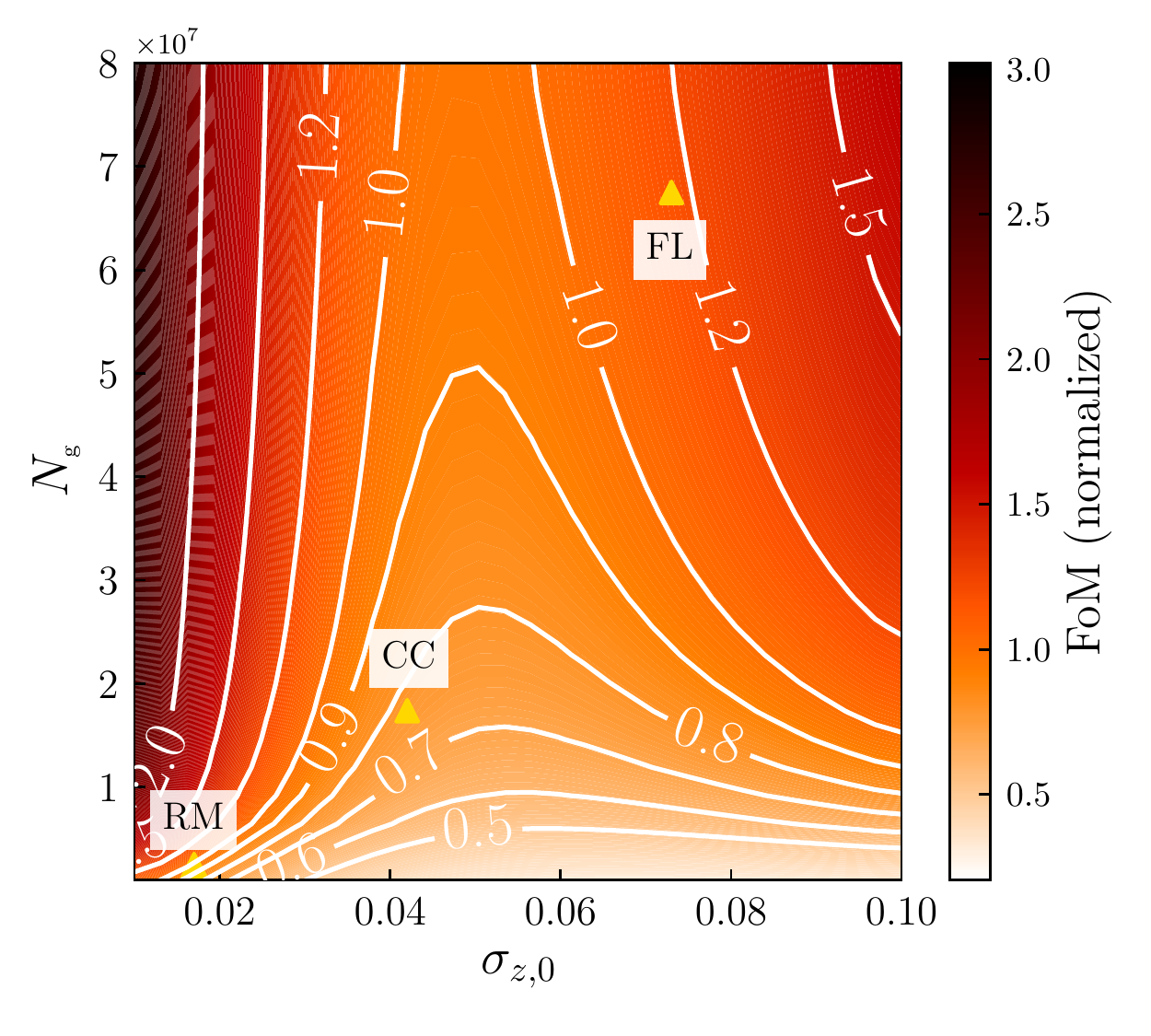}}
\subfigure[]{\includegraphics[width=\columnwidth]{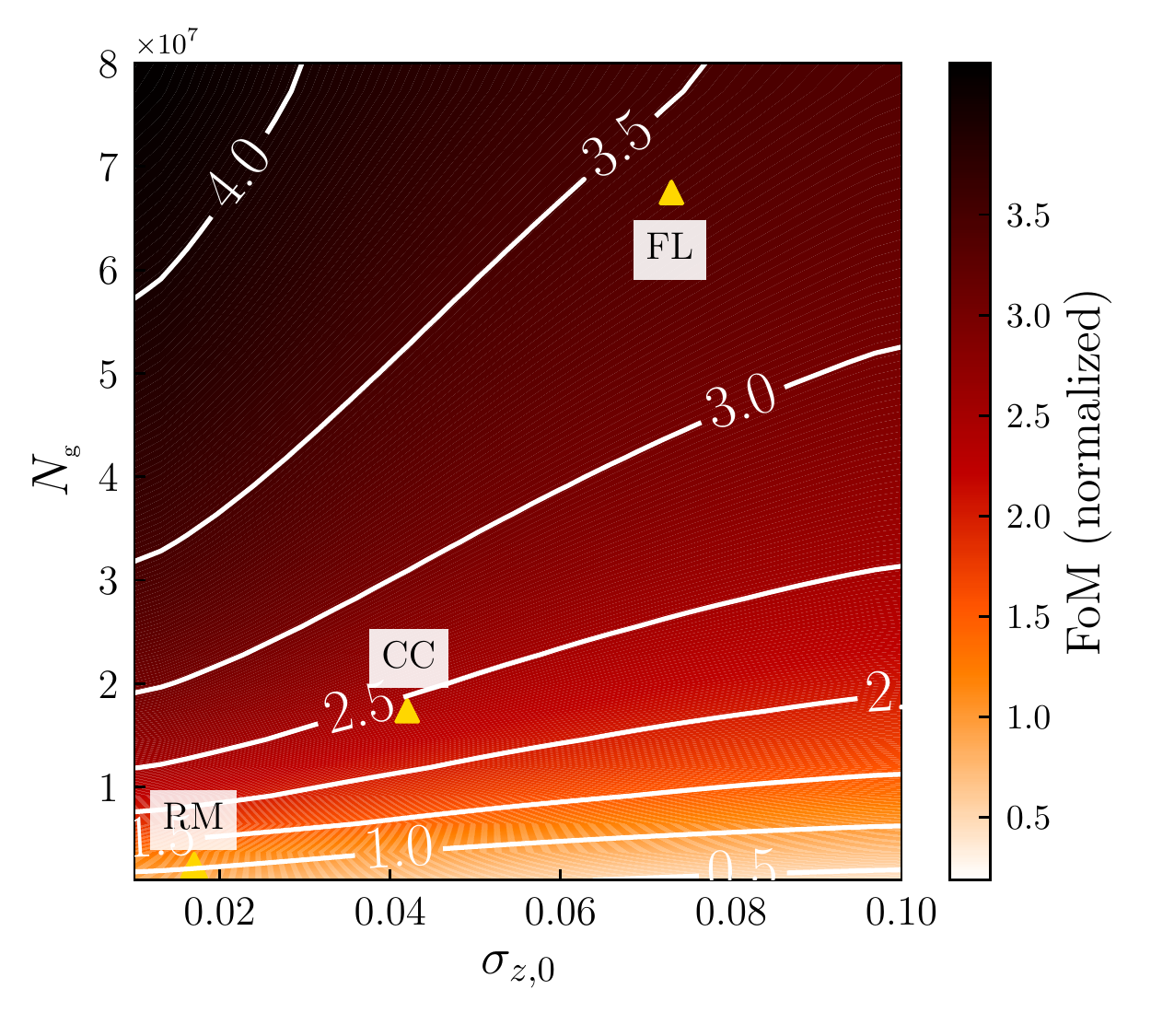}}
\caption{The figure of merit (FoM) for the set of parameters $\Omega_m - \sigma_8$ as a function of the photo-z scatter, $\sigma_{z,0}$ and galaxy sample size, $N_{\mbox{\scriptsize{g}}}$. The four figures correspond to different assumptions about our prior knowledge of the photo-z parameters: (a) Flat priors are assumed, photo-z parameters totally free. (b) A conservative prior of the form $\sigma(\sigma_{z,0}) = \sigma(z_{\mbox{\scriptsize{b}}}) = 0.4\sigma_{z,0}$ is assumed. (c) An optimistic prior of the form $\sigma(\sigma_{z,0}) = \sigma(z_{\mbox{\scriptsize{b}}}) = 0.04\sigma_{z,0}$ is assumed. (d) The photo-z parameters are held fixed to their fiducial values. In all cases the results are normalized to the FoM of the redMaGiC sample for the specific case, so only the relative differences between samples are shown. The overall FoM increases as we tighten our photo-z  priors (see main text).}
\label{fig: Baseline}
\end{figure*}

\section{Baseline comparison}
\label{sec: Baseline}

We now proceed to study the constraining power from a range of different samples beyond the redMaGiC, flux-limited and color cuts-defined samples. In all cases, we consider angular clustering in the standard five redshift bins $z \in [0.2, 0.95]$, without taking into account the cross-correlations, as we did in our earlier example.

\subsection{The photo-z scatter -- sample size space}
\label{subsec: sig_sample_space}

 We consider a grid in photo-z scatter, $\sigma_{z,0}$, and galaxy sample size, $N_{\mbox{\scriptsize{g}}}$, space, with range:
\begin{eqnarray}
0.01\leq &\sigma_{z,0} &\leq 0.1 \\
1.0 \times 10^6 \leq &N_{\mbox{\scriptsize{g}}} &\leq 8.0 \times 10^7.
\end{eqnarray}
With these choices we cover a wide range of different samples, with the redMaGiC and flux-limited ones being two extreme cases in the photo-z precision  -- sample size parameter space.

In order to compare different samples quantitatively, we use the figure of merit (FoM) defined in Eq. \eqref{eq: FoM_1}, computed for the pair of cosmological parameters $\mathbf{\theta} =[ \Omega_m, \sigma_8]$ in the context of $\Lambda$CDM model, keeping the other cosmological parameters fixed to their fiducial values and marginalizing over the photo-z parameters. For the cosmological parameters we use the fiducial parameters stated in the introduction. With this choice we can characterize the constraining power of different samples using a single number which, in our case, is proportional to the inverse of the area of the confidence ellipses like those presented in Fig. \ref{fig: Data_and_model}; a higher FoM corresponds to tighter constraints.

In Fig. \ref{fig: Baseline} we present the FoM in the grid of samples defined above for four different choices about the priors imposed on the photo-z parameters. To make the comparison between samples easier, we normalize the value of the FoM to that of the redMaGiC sample in each case, so we actually plot FoM/FoM$_{\mbox{\scriptsize{RM}}}$. We also overplot (yellow triangles) the positions of the redMaGiC (RM) flux-limited (FL) and color cuts-defined (CC) samples in the $\sigma_{z,0} - N_{\mbox{\scriptsize{g}}} $ plane. 

\begin{figure}
\centering
\includegraphics[width=0.8\columnwidth]{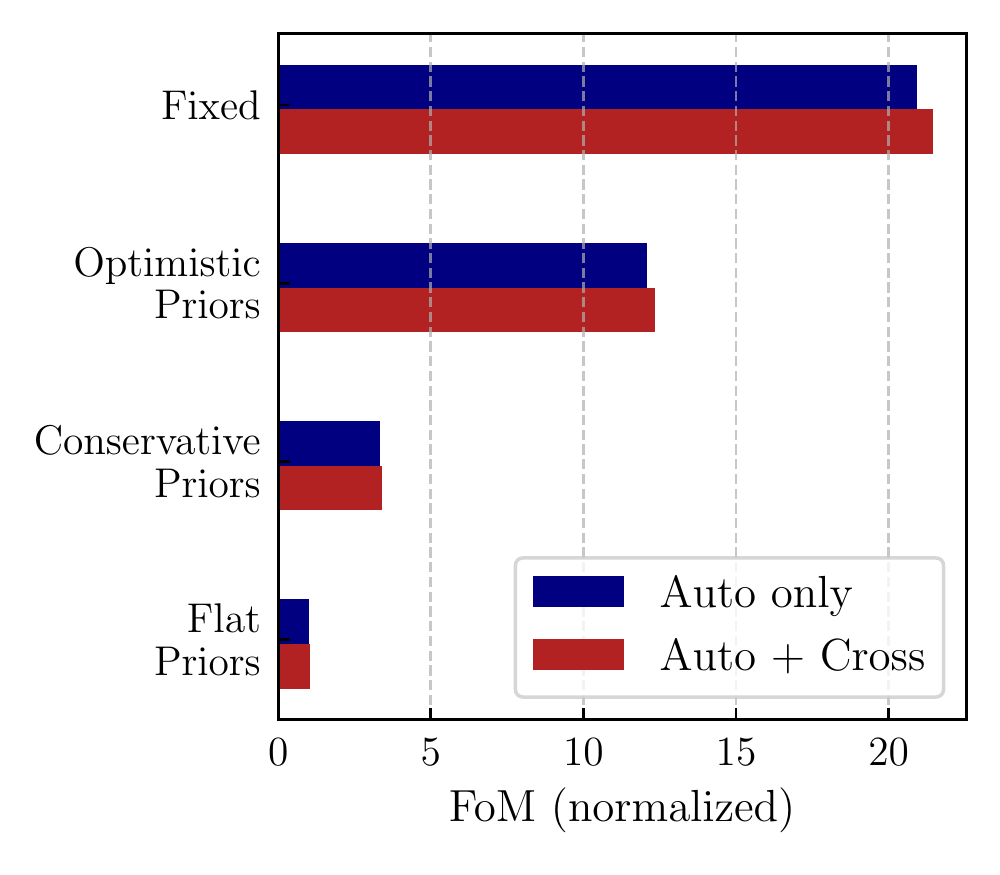} 
\caption{Increase of the FoM of the RM sample as we tighten our photo-z parameter priors: by introducing conservative priors we get a $\sim$ 3 times higher FoM than having flat, uninformative, priors (the case where we set the FoM to be equal to one). This increases to a $\sim$ 12 times higher FoM when optimistic priors are assumed and $\sim$ 21 times higher FoM if we assume a perfect knowledge of photo-zs. The inclusion of cross-correlations does not change the results significantly, since for the RM sample the overlap between redshift bins is minimal.}
\label{fig: Norm_increase}
\end{figure}

In panel (a), of Fig. \ref{fig: Baseline} we assume flat priors on the photo-z parameters; we leave them free to vary and marginalize over them. We see that the FoM is higher for smaller values of the photo-z scatter and larger sample sizes. The FL sample gives a FoM that is $\sim 1.62$ higher than that of the RM sample, while the CC sample performs slightly better, giving a FoM that is $\sim 1.81$ higher that that of the RM sample.

The introduction of priors on the photo-z parameters affects the results in two ways: first by changing the relative FoM between different samples and second by an overall shift of the FoM; tighter priors result in better constraints for all samples (see Fig. \ref{fig: Norm_increase}). In panels (b) and (c) we adopt Gaussian priors of the form $\sigma(\sigma_{z,0}) = \sigma(z_{\mbox{\scriptsize{b}}}) = 0.4(0.04)\sigma_{z,0}$, respectively. We assume that the priors simply scale with the photo-z scatter; this is a good approximation for photometric redshifts externally calibrated using spectroscopic samples. For simplicity we assumed that the constant of proportionality is the same for all samples; in the case of panel (b) this corresponds to conservative priors, comparable to those used in the DES Y1 analysis; those of panel (c) correspond to optimistic priors, similar to those used in forecasts for the Large Synoptic Survey Telescope (LSST) {\footnote{\url{https://www.lsst.org/}}} project (e.g., \citealt{Schaan2017}).

The conservative priors lead to an overall increase of the FoM by a factor of $\sim 3$, while using the optimistic priors there is an overall increase by a factor of $\sim 12$, compared to the flat prior case (see also Fig. \ref{fig: Norm_increase}). Furthermore, as expected, the relative FoM of samples changes as well; for example, in the conservative priors case, the FL sample gives a FoM $\sim 1.47$ times higher than the RM sample, while in the optimistic priors case the FL sample gives a FoM $\sim 1.16$ times higher than the RM one. In both cases the difference is lower than in case (a) where flat priors were assumed. This is because of the assumption of the priors being proportional to the scatter: samples with smaller photo-z scatter are less constrained as well.

Finally, in panel (d) we show the case where the photo-z parameters are kept fixed in their fiducial values; in other words a perfect knowledge of them is assumed. This case represents the maximum information we can get from a particular sample. Under this assumption we have an overall increase of the FoM ($\sim 21$ times higher than the flat prior case); also now the FoM of the FL sample is $\sim 3.40$ higher than that of the RM sample and that of the CC sample is $\sim 2.43$ times higher.

Generally, in all cases, we see that better photo-zs and larger samples are beneficial. However, there are some specific points worth noticing. For example, we observe that in most case the increase of the FoM with sample size is much more rapid in the region where photo-zs are better; this suggests that even a small increase in high-quality photo-zs can be extremely beneficial. Interestingly, we find that in most cases some significantly larger samples with smaller photo-z scatter have potentially better constraining power than the redMaGiC sample.

We also see that the specific level of priors can introduce non-trivial trade-offs as in case (c). We would like to discuss this case in more detail, since its behavior differs significantly from that of the other panels, in the sense that in some cases it seems that samples with the same size but larger photo-z scatter give better constraints. We have checked that this is a result of the particular level of (relative) photo-z priors used in that case; we have found that moderate-to-tight priors improve the FoM for samples with larger $\sigma_{z,0}$ more than they do for samples with small $\sigma_{z,0}$. As an example we considered the behavior of two samples, both with size $N_{\mbox{\scriptsize{g}}} = 5 \times 10^7$ and photo-z scatter $\sigma_{z,0} = 0.05$ and $\sigma_{z,0} = 0.08$, respectively. Defining  $r$ as the ratio $\sigma(\sigma_{z,0})/\sigma_{z,0}$ of the photo-z scatter prior to the value of the scatter itself, we find that the sample with lower photo-z scatter gives a higher FoM in the limit of weak ($r \gtrsim 0.2$) or very informative ($ r \lesssim 0.008$) priors. However, in the intermediate priors range ($ 0.008 \lesssim r \lesssim 0.2$) the sample with higher photo-z scatter gives better constraints.

To better understand this behavior, and why constraints flip their order in the intermediate priors case, we consider a toy model where  our Fisher analysis includes only one cosmological parameter (say, the matter density, $\Omega_m$) and one photo-z parameter (the scatter $\sigma_{z,0}$) with prior $\sigma_p \equiv \sigma(\sigma_{z,0})$. In the limit $\sigma_p \ll 1$, the forecast constraints on $\Omega_m$ can be expressed in terms of the diagonal terms, $F_{11}$ and $F_{22}$, of the $2 \times 2$ Fisher matrix (see the definition, Eq. \eqref{Eq: Fisher_mat}, with $1 \equiv \Omega_m$ and $2 \equiv \sigma_{z,0}$) , and the prior $\sigma_p$, as $\sigma^2_{\Omega_m} \cong (1 + \sigma_p^2 F_{22})/F_{11}$.  Both $F_{11}$ and $F_{22}$ decrease with increasing $\sigma_{z,0}$. When $\sigma_p \to 0$ then $\sigma_{\Omega_m} \cong 1/F_{11}$ and thus we get tighter cosmological constraints for the sample with larger $F_{11}$, which is the one with smaller $\sigma_{z,0}$. When non-zero priors are introduced, the second term,  
 $\sigma_p^2F_{22}/F_{11}$, cannot be neglected and the tighter cosmological constraints come from the sample for which this term is smaller. The dependence of this term for fixed $N_{\mbox{\scriptsize{g}}}$  and increasing $\sigma_{z,0}$ depends not only on $\sigma_{z,0}$ itself but also the ratio of $F_{22}/F_{11}$, which is not trivially dependent on $\sigma_{z,0}$. Numerically, we find that as $\sigma_{z,0}$ increases, it is possible for this second term to either increase or decrease. Although we discussed here only a toy example, the same argument can be used to understand the non-intuitive behavior in case (c) of Fig \ref{fig: Baseline}.

In the above paragraphs we studied the behavior of the FoM as a function of $\sigma_{z,0}$ and $N_{\mbox{\scriptsize{g}}}$, covering a wide range of samples. However, we note that not all regions in the plot are allowed. For example, with the current photometric redshift estimation techniques,  accurate photometric redshifts can be estimated only for a small fraction  of the surveyed galaxies. Roughly, samples selected from photometric survey data are expected to lie close to the band that connects the triangles denoting the RM, CC and FL samples, with the exact position depending on the details of the selection cuts.

\subsection{Dependence on priors}
\label{subsec: Depend_on_priors}

\begin{figure*}
\centering
\begin{multicols}{2}
\includegraphics[width=\linewidth]{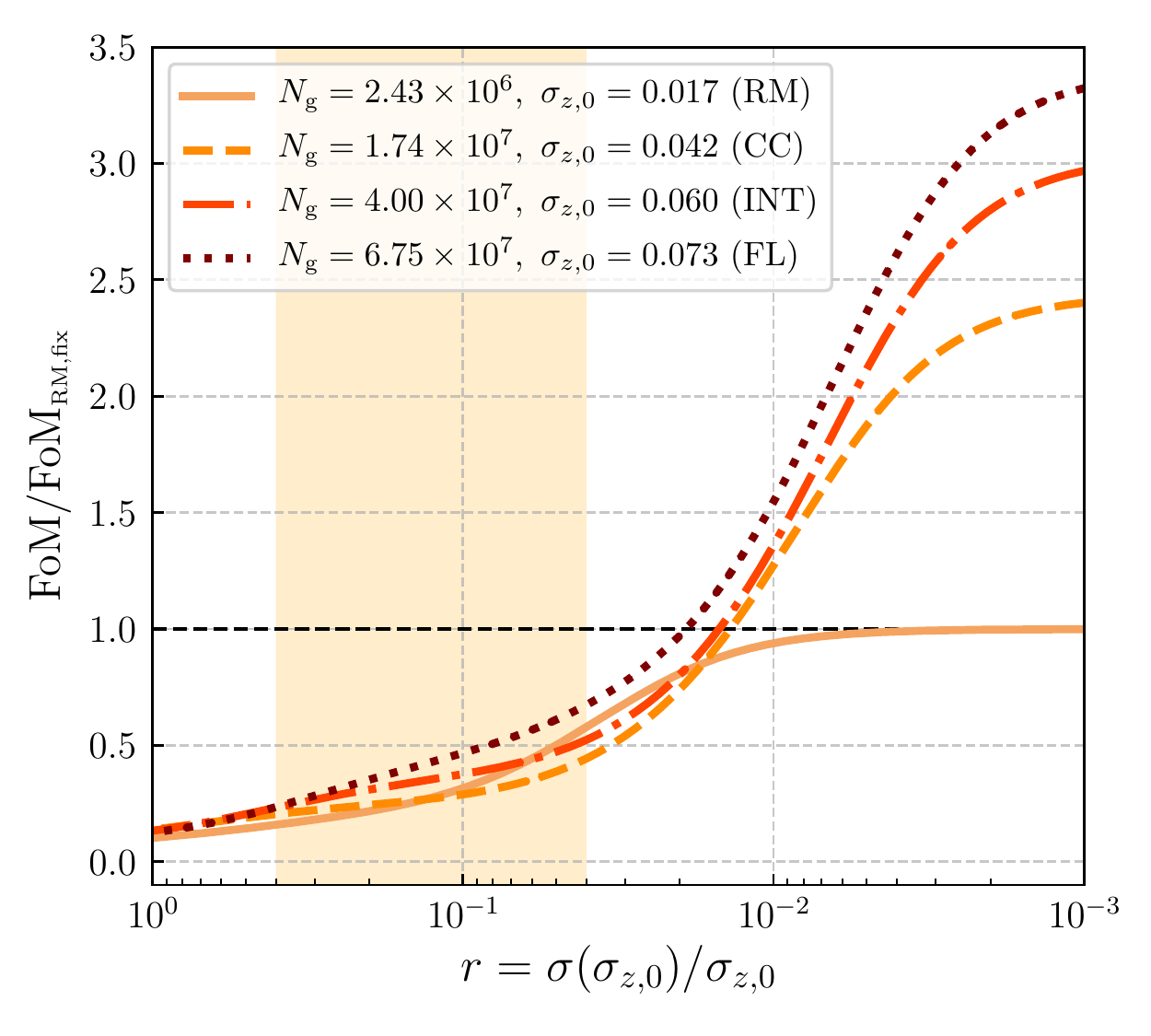}\par 
\includegraphics[width=\linewidth]{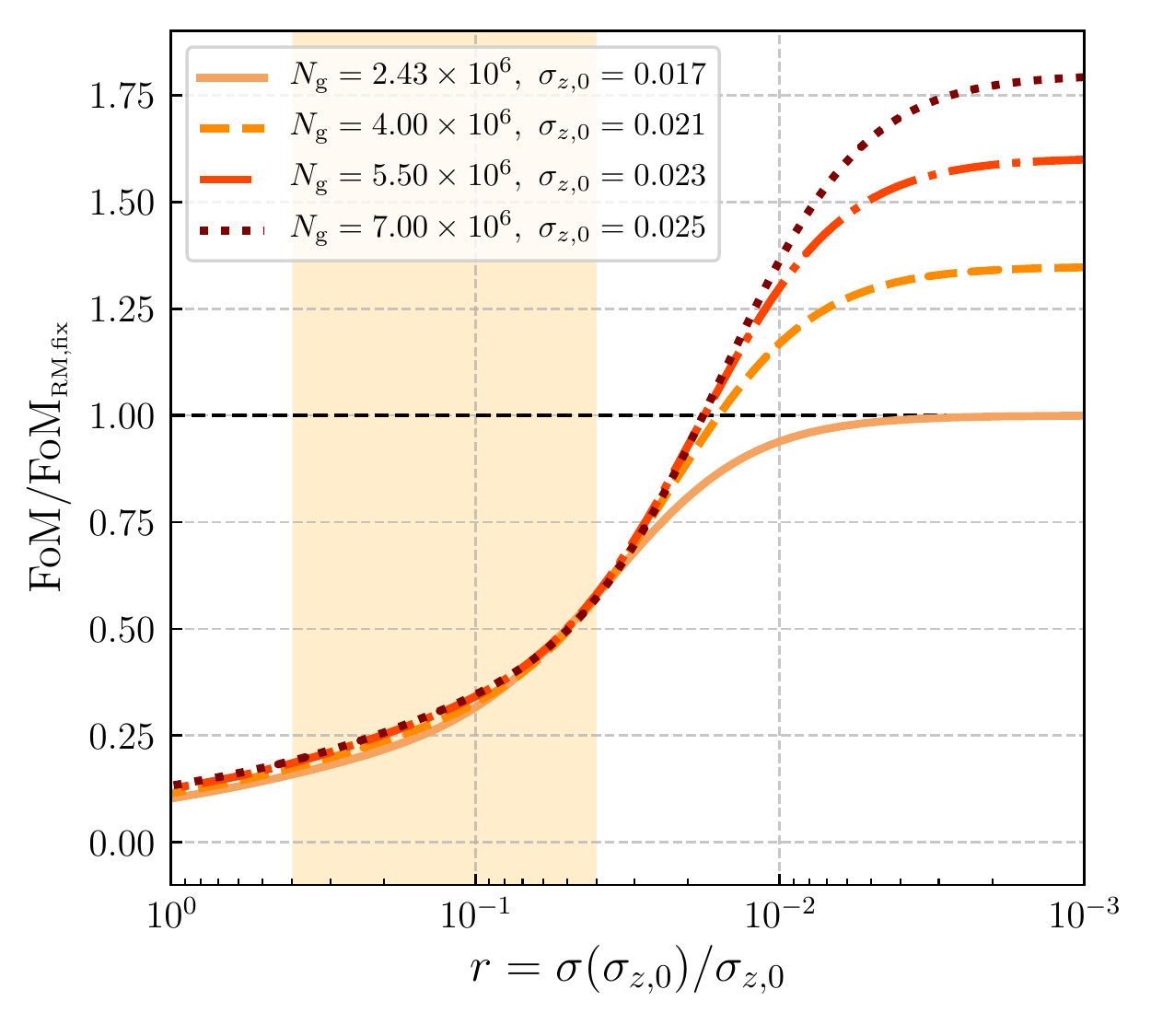}\par 
\end{multicols}
\caption{The FoM as a function of the ratio of the width of the prior to the photo-z scatter, $\sigma(\sigma_{z,0})/\sigma_{z,0}$, for some characteristic samples. The FoM is normalized to that of the redMagiC sample in the  case where we fix the photo-z parameters to their fiducial values. The ratio $r$ is a rough measure of the calibration requirements for a sample to achieve a certain FoM. \textit{Left panel:} Four samples between the redMaGiC and flux-limited samples. \textit{Right panel:} Four samples close to redMaGiC. The yellow bands show the plausible prior range $0.04 < r < 0.4$ (see main text).}
\label{fig: Priors_baseline}
\end{figure*}

We have seen that the level of uncertainty on the photo-z parameters, quantified through the priors on them, significantly affects the cosmological constraints from galaxy clustering. Thus, to compare the constraining power of different samples, the photo-z priors have to be carefully taken into account.

In the previous section we explored the effects of introducing two different types (conservative and optimistic) of priors on the photo-z parameters of the samples. One simplifying assumption made there was that the priors on both the photo-z scatter and bias were proportional to $\sigma_{z,0}$, with the constant of proportionality being the same for all samples. Furthermore, as we said, just two such values were considered. Here, we examine how different priors affect our results, by studying the behavior of the FoM  as a function of the photo-z parameter priors, for some characteristic samples.

The priors on the photo-z parameters of samples used for cosmological analyses come from external calibration; the most direct method is to use a representative sub-sample for which spectroscopic redshifts or high-precision photometric redshifts are also available. If the size of the calibration sample is $N_{\mbox{\scriptsize{cal}}}$, then a simple model for the uncertainties of the photo-z parameters is \citep{Ma2006, Hearin2010, Newman2015}:
\begin{equation}
\label{eq: simple_prior_model_1}
\sigma(\sigma_z) = \sigma_z \sqrt{\frac{1}{2N_{\mbox{\scriptsize{cal}}}}},
\end{equation}
\begin{equation}
\label{eq: simple_prior_model_2}
\sigma(z_{\mbox{\scriptsize{b}}}) = \frac{\sigma_z}{\sqrt{N_{\mbox{\scriptsize{cal}}}}}.
\end{equation}
In practice, photometric surveys use other methods as well for the calibration of photometric samples (for example cross-correlations with high-precision redshift samples, e.g., \cite{Cawthon2018, Gatti2018}). Furthermore, Eqs. \eqref{eq: simple_prior_model_1}-\eqref{eq: simple_prior_model_2} are optimistic about the calibration requirements: they assume Gaussian photo-zs, absence of catastrophic outliers and that the calibration sample is a fair representation of the photometric sample.

Here we assume, as in Section \ref{subsec: sig_sample_space}, that priors are proportional to the photo-z scatter, so we write $\sigma(\sigma_{z,0}) = \sigma(z_{\mbox{\scriptsize{b}}}) = r  \sigma_{z,0}$, with $r \equiv \sigma(\sigma_{z,0})/\sigma_{z,0}$  being the proportionality constant, that quantifies the calibration needs of a sample, to achieve a certain prior. Intuitively, it makes sense to assume that samples with higher photo-z scatter are also harder to calibrate.

In Fig. \ref{fig: Priors_baseline} we plot the FoM as a function of the parameter $r$ for a number of samples. In the left-hand side panel we plot the RM, FL and CC, with one more additional sample that sits between CC and FL. We have chosen to plot the horizontal axis in inverted order (higher to lower); lower values of $r$ mean more constrained samples.

We also plot the RM sample and three other small variants of RM in the right-hand side panel of Fig.  \ref{fig: Priors_baseline}. These variants are samples with slightly smaller $\sigma_{z,0}$ and larger sizes than the RM sample, and can be obtained by applying magnitude and color cuts, in a way similar to those used in Section \ref{subsubsec: Color_cuts}. We normalize the FoM to the FoM of redMaGiC when fixed photo-z parameters are assumed, FoM$_{\mbox{\scriptsize{RM,fix}}}$.

Looking at the left-hand side panel of Fig. \ref{fig: Priors_baseline},  we see that for values of the parameter $r > 0.01$ all samples have generally comparable FoMs. Interestingly, the flux-limited sample ($N_{\mbox{\scriptsize{g}}} = 6.75 \times 10^7, \,\, \sigma_{z,0} = 0.073$) seems to always give better results than the redMaGiC sample ($N_{\mbox{\scriptsize{g}}} = 2.43 \times 10^6, \,\, \sigma_{z,0} = 0.017$), with significantly higher FoM for well-calibrated samples ($r<0.01$). Given that even in optimistic forecasts, for future DES and LSST analyses we expect $r > 0.04$ (see e.g., \citealt{Newman2015,Schaan2017,LSST2018,Buchs2019}), we conclude that there is a potential gain in using the FL sample, but  not the other samples since they give lower FoM than the RM one when $0.04 < r < 0.1$.

Similarly, for the samples that are extensions of the redMaGiC one, the gains from using an alternative sample are not significant in the plausible prior range. For  $r > 0.04$ all samples seem to give comparable constraints. Only for $ r < 0.04$ they start outperforming the redMaGiC sample, with  a FoM $\sim 1.5$ larger than the redMaGiC for $r = 0.01$. 

The above analysis suggests that in the scenario presented in in this section (clustering in five bins of width $\delta z = 0.15$, without cross-correlations) there can be a substantial ($\sim 20-50\%$) gain from considering  much larger samples than the RM, like the FL sample. However, we should be cautious in our conclusions, since for such samples the Gaussian approach taken here is rather simplistic and ignores effects such as catastrophic photo-z outliers, and the small benefits of using a FL sample presented in Fig. \ref{fig: Baseline} may not hold in a more detailed analysis. 

Overall, our analysis in this and in the previous section (the counter-intuitive behavior of Fig. (\ref{fig: Baseline}
c)) suggests that the resulting FoM is extremely sensitive to the choice of priors, and thus they should be carefully considered in practical applications of sample selection.

\section{Including cross-correlations}
\label{sec: Cross_correlations}

\begin{figure*}
\centering
\subfigure[]{\includegraphics[width=\columnwidth]{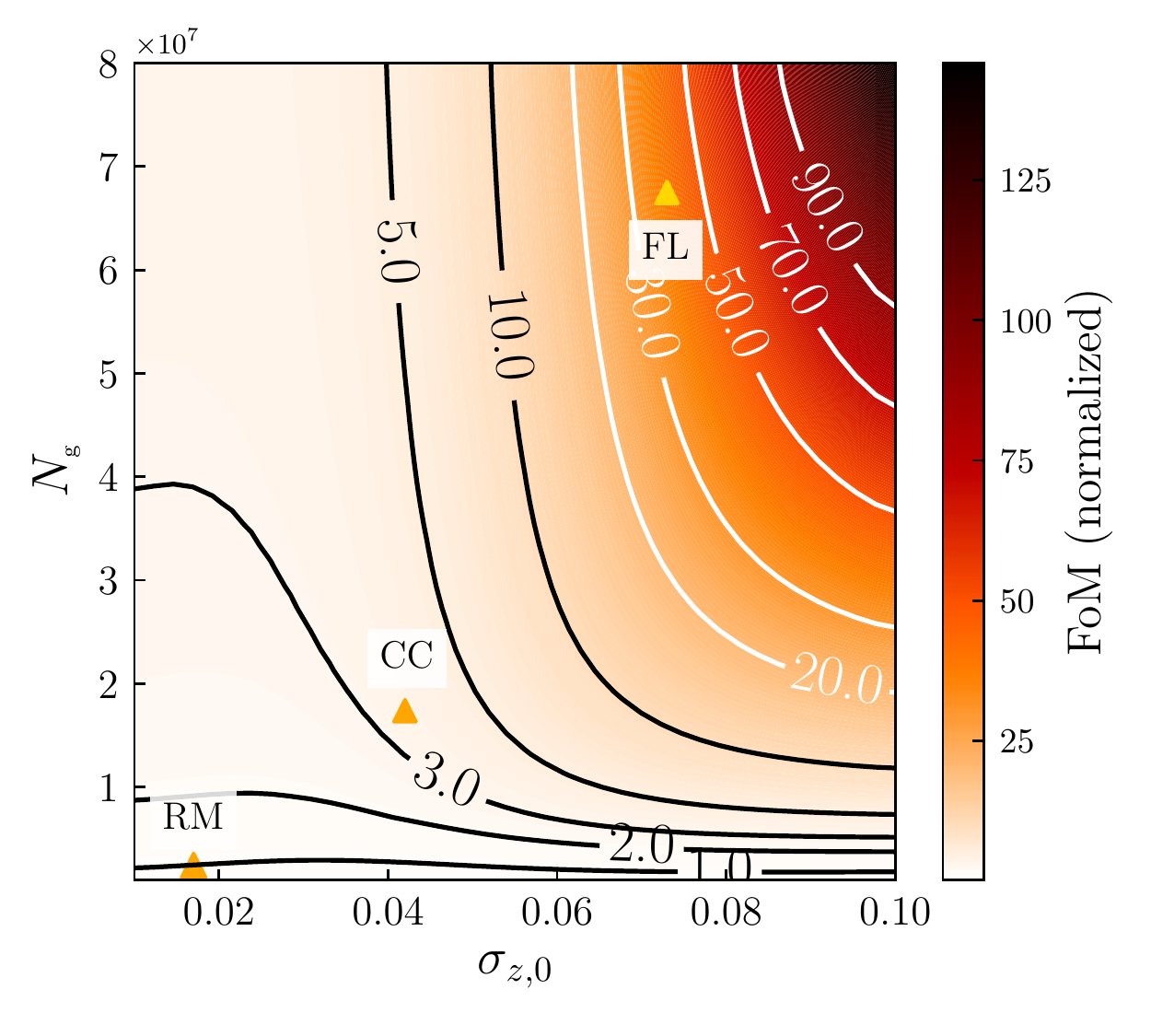}}
\subfigure[]{\includegraphics[width=\columnwidth]{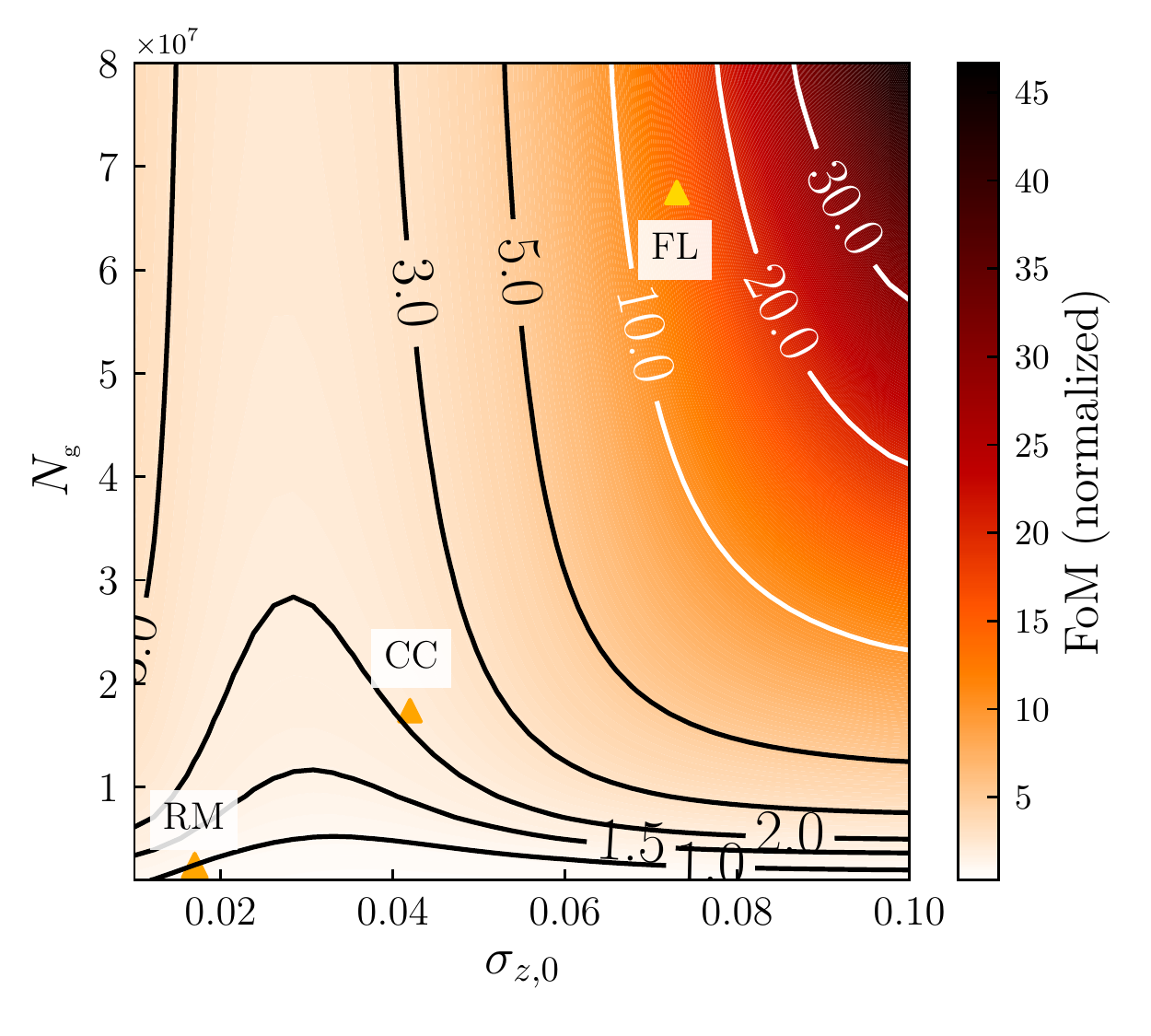}}

\subfigure[]{\includegraphics[width=\columnwidth]{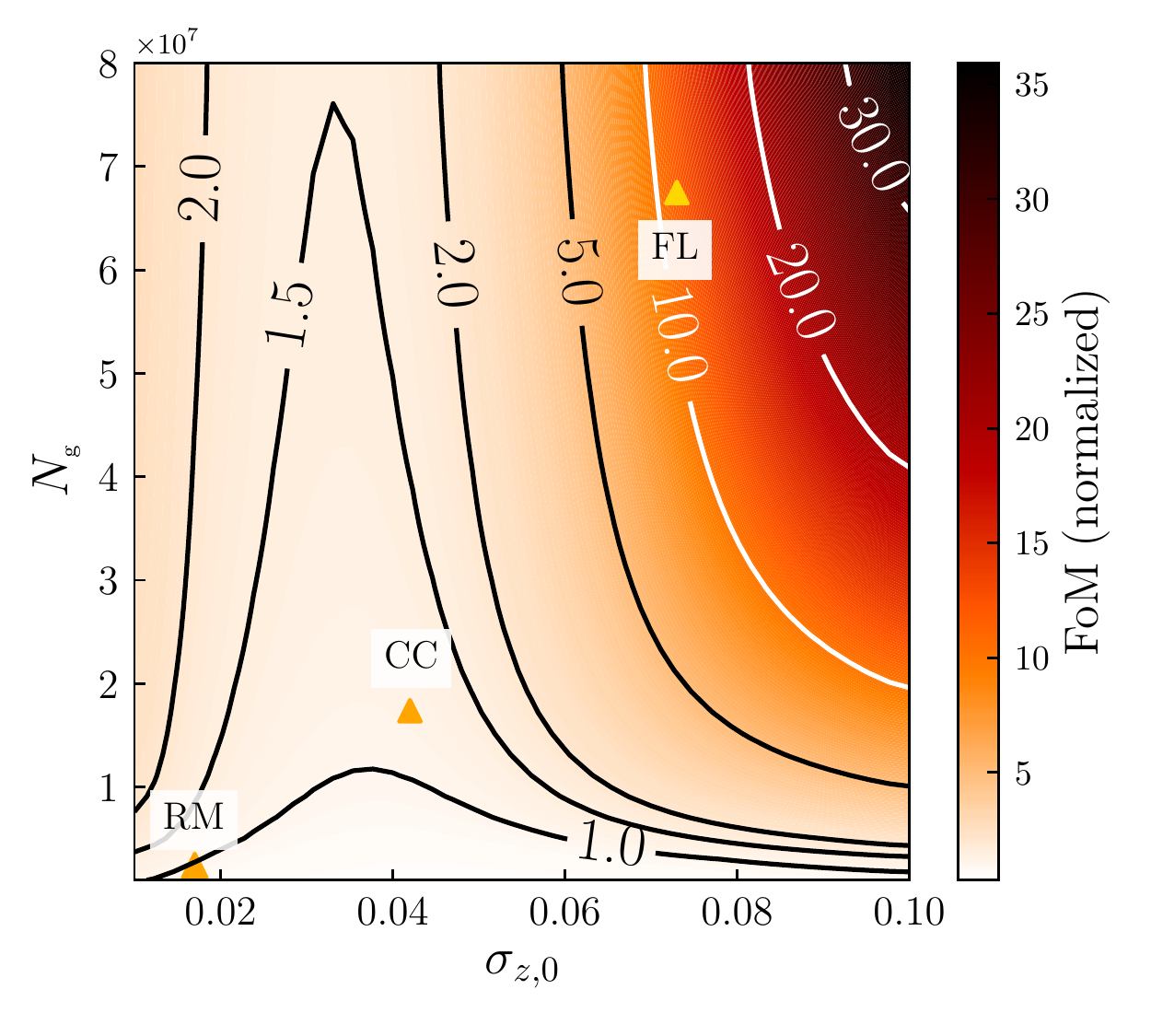}}
\subfigure[]{\includegraphics[width=\columnwidth]{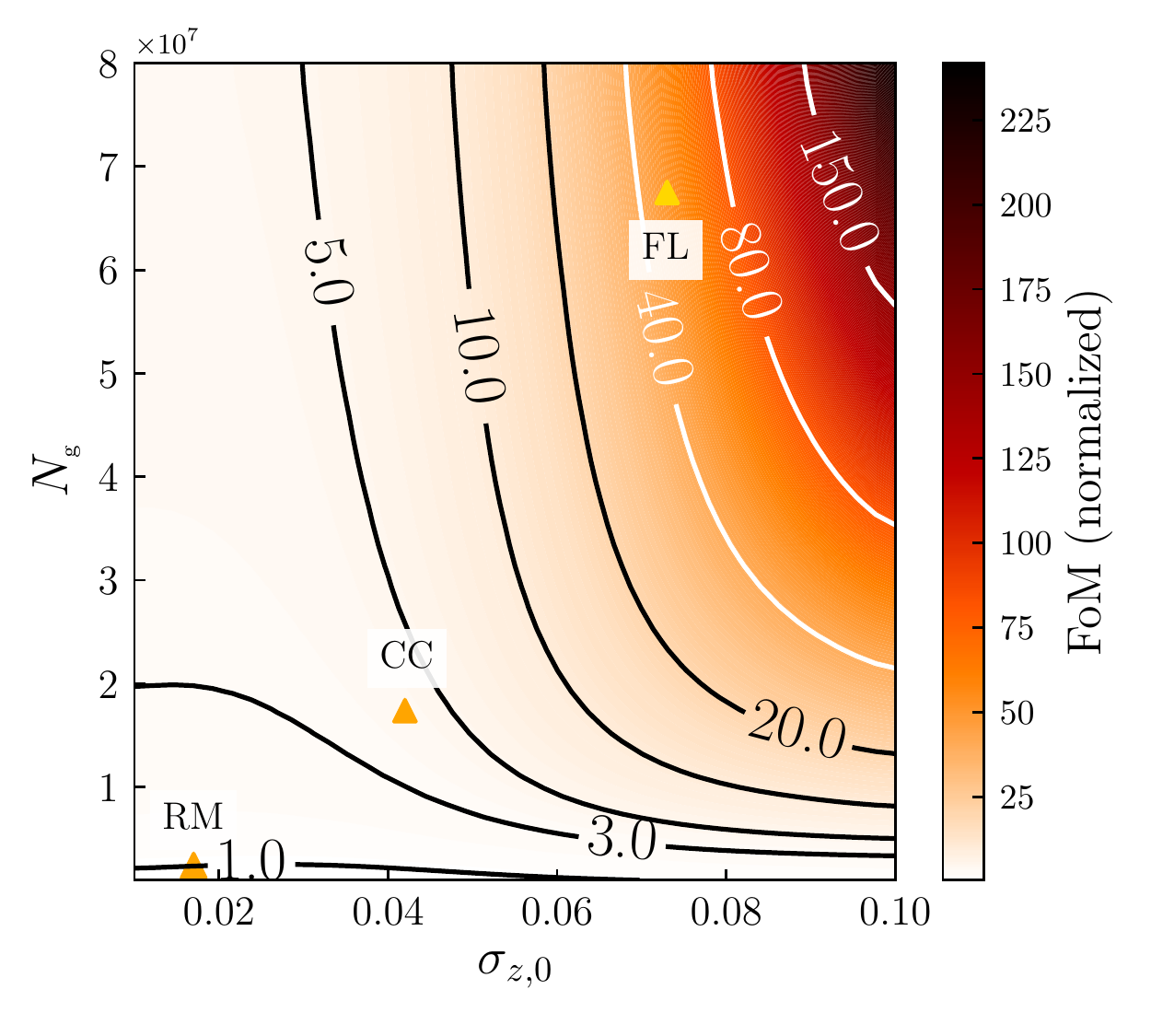}}
\caption{Similar to Fig. \ref{fig: Baseline}, but now including the cross-correlations between bins when forecasting the cosmological constraints from different samples.}
\label{fig: With_cross}
\end{figure*}

So far we have ignored the cross-correlations between redshift bins. Such an approach, that was used, for example, in the first year of DES clustering analysis makes sense when samples with accurate (scatter much smaller than the redshift bin width) photo-zs are used. In that case, the overlap between redshift bins is minimal and thus the information from the cross-correlation spectra is not significant.

If we want to explore the possibility of using larger samples with higher photo-z uncertainties (and thus overlap between bins) in future analyses, we have to examine the importance of including the cross-correlations. In this section we study the cosmological constraints from a combined analysis of auto and cross spectra for a range of different samples.

\begin{figure}
\centering
\subfigure[]{\includegraphics[width=0.46\columnwidth]{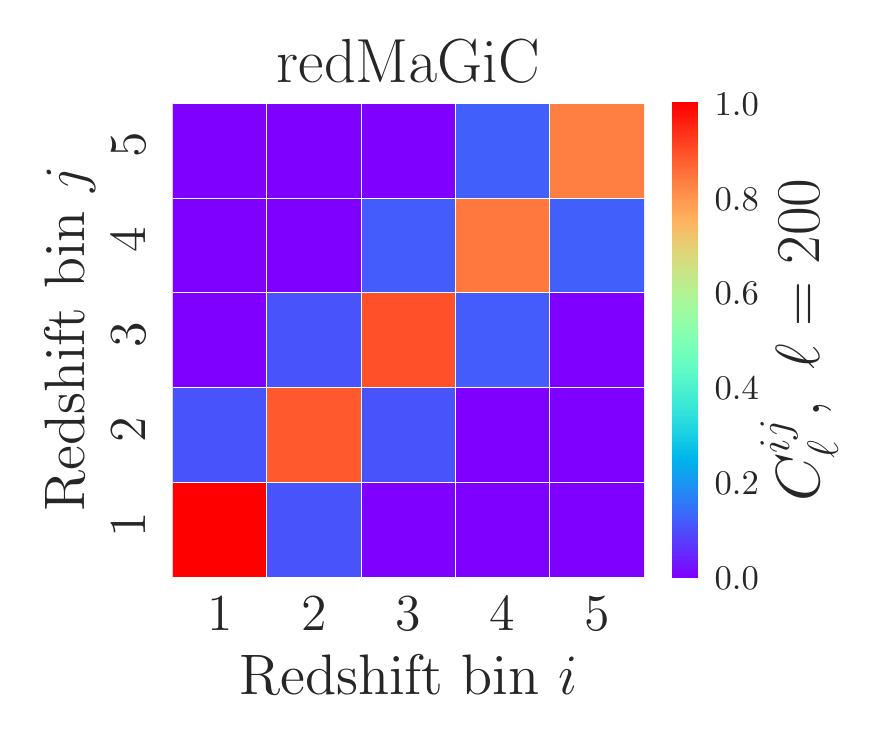}}
\subfigure[]{\includegraphics[width=0.46\columnwidth]{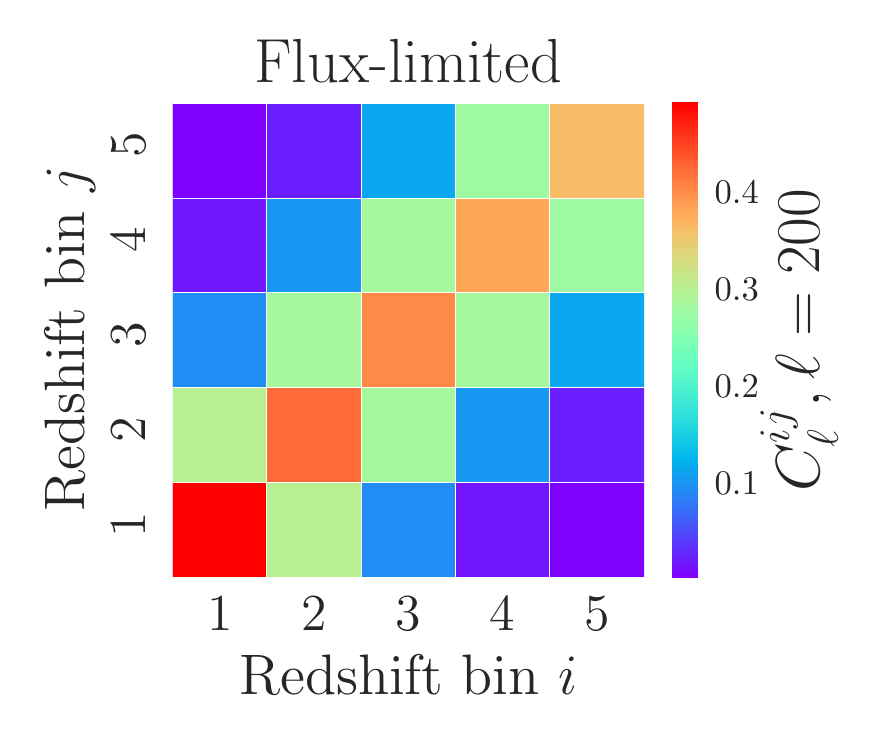}}
\caption{The value of the angular auto- and cross-spectra, at $\ell =200$ for (a) the redMaGiC and (b) the flux-limited samples. Because of higher value of photo-z scatter, the overlap between redshift bins is higher for the flux-limited sample, and so is the value of the cross-spectra.}
\label{fig: Importance}
\end{figure}

In Fig. \ref{fig: With_cross} we present the FoM as a function of the photo-z scatter, $\sigma_{z,0}$,  and sample size, $N_{\mbox{\scriptsize{g}}}$, for the same range of values as in Section \ref{sec: Baseline}. As in that section, the four panels  correspond to four different photo-z prior choices: (a) flat priors on the photo-z parameters; (b) Gaussian priors of the form $\sigma(\sigma_{z,0}) = \sigma(z_{\mbox{\scriptsize{b}}}) = 0.4  \sigma_{z,0}$ (conservative); (c) Gaussian priors of the form $\sigma(\sigma_{z,0}) = \sigma(z_{\mbox{\scriptsize{b}}}) = 0.04  \sigma_{z,0}$ (optimistic)  and (d) fixed photo-z parameters.

Again, as in Section \ref{sec: Baseline}, the FoM in each panel is normalized by dividing by the FoM of the redMaGiC sample. In Fig. \ref{fig: Norm_increase}  we show how the FoM of the redMaGiC sample (the normalization constant) improves with tighter photo-z priors (red bars). The improvement is similar to the auto-spectra only case ($\sim 3$ times to $\sim 12$ times to $\sim 21$ times improvement as we go from case (a) to (d)). Also, note that the FoM of the redMaGiC sample improves just slightly with the inclusion of the cross-correlations. This confirms our previous notion that for accurate photo-z samples (as the redMaGiC) the additional information from the cross-spectra is negligible.

On the other hand, the improvement of the FoM of large samples with high photo-z scatter is very significant. For example, in all four cases presented in Fig. \ref{fig: With_cross}  the flux-limited sample has a FoM that is over an order of magnitude higher than that of the redMaGiC. This is in contrast to the case with only the auto-spectra, where the gain by going from the RM to the FL sample is only marginal. 

The behavior of the FoM in Fig.  \ref{fig: With_cross} may seem to be counter-intuitive. Compare again these plots with those of Fig. \ref{fig: Baseline}. In the auto-spectra only case, the FoM peaks at the limit of large, most accurate photo-z samples (upper left corner of the plots). The FL sample may (slightly) outperform the RM one, but only because its size is large enough to compensate for the loss cause by the less accurate photo-zs. However, when cross-correlations are included the FoM peaks at the region of large samples and less accurate photo-zs (upper right corner of the plots). The counter-intuitive part is that when the cross-correlations are taken into account, it seems to be preferable to use  samples with larger $\sigma_{z,0}$.

As we will explain in detail in the following section, what the above really suggests is that with our binning choice, for samples with the same size, we do not take full advantage of the more accurate photo-zs. In our discussion so far we have considered angular galaxy clustering in the same five bins of constant width $\delta z = 0.15$ for all samples. For such a bin width, the overlap between the true redshift distributions of adjacent  redshift bins is significant when samples with high photo-z uncertainties ($\sigma_{z,0} \sim 0.07-0.1$), like that of the FL sample, are considered. Thus, significant information can be  obtained from the cross-correlation spectra. As we are moving to samples with more accurate photo-zs, we gain information from the auto-spectra that increase in amplitude (see the relevant discussion in Section \ref{sec: Samples_Constraints}). At the same time there is a significant drop of the overlap between redshift bins and the information from the cross-spectra. For wide enough bins, the gain we get from the auto spectra is not enough to compensate the loss from the cross spectra (for wide bins we have erased the radial information anyway), leading to the behavior we see in Fig. \ref{fig: With_cross}.

To put the above discussion into  more quantitative perspective, we plot in Fig. \ref{fig: Importance} the value of the angular auto and cross spectra at a characteristic scale of $\ell = 200$. As expected, the RM sample gives higher values in the auto spectra but all cross-correlations are almost zero. The FL sample has lower auto-spectra values, consistent with our previous discussion,  but now the cross-spectra are very significant, having comparable values to those of the auto-spectra. This, combined  with the fact that noise in the auto-spectra is significantly higher for the RM sample than the FL one, explains how the FL sample results in a significantly higher FoM.

\begin{table}
\centering
\caption{Self-calibration of the photo-z scatter, $\sigma_{z,0}$, and photo-z bias in the second bin, $z^{(2)}_{\mbox{\scriptsize{b}}}$, parameters of two samples with the same size, $N_{\mbox{\scriptsize{g}}} = 5 \times 10^7$, and scatter $\sigma_{z,0} = 0.02, 0.08$ when cross-correlations: (i) are included, (ii) are not included.}
 \label{tab: Self_cal}
 \begin{tabular}{lcc}
  \hhline{===}
   &   {\textbf{(i) With Cross-Correlations}}&  \\
  \hline
   &  $\sigma(\sigma_{z,0})$    &  $\sigma(z^{(2)}_{\mbox{\scriptsize{b}}})$ \\
  \hline
 $ \sigma_{z,0} = 0.02$ & 0.0023 & 0.015 \\
  $\sigma_{z,0} = 0.08$ & 0.0014 & 0.005 \\
  \hline
  \hhline{===}
  & {\textbf{(ii) Without Cross-Correlations}} &  \\
   \hline
  &   $\sigma(\sigma_{z,0})$   &  $\sigma(z^{(2)}_{\mbox{\scriptsize{b}}})$  \\
  \hline
  $\sigma_{z,0} = 0.02$ & 0.0024 & 0.016 \\
  $\sigma_{z,0} = 0.08$ & 0.0071 & 0.031 \\
  \hline
 \end{tabular}
\end{table}

A similar finding, that overlapping bins can significantly improve cosmological constraints, has bin presented in \citet{Nicola2014} for the case where bin overlap can be modeled as a Gaussian redshift bin broadening (this corresponds to perfectly known Gaussian photo-z parameters  in our case). We find that this conclusion holds even in the case of very weak prior knowledge of the photo-z parameters, since the inclusion of cross-correlations leads to a significant self-calibration of the photo-z parameters for higher $\sigma_{z,0}$ (more bin overlap).

By self-calibration, we mean constraints on the photo-z parameters from angular clustering measurement itself, without invoking external calibration coming from any of the methods mentioned in Section \ref{subsec: Depend_on_priors}.

In Table \ref{tab: Self_cal} we present the self-calibration level of the photo-z scatter, $\sigma(\sigma_{z,0})$, and bias in the second redshift bin (as an example, we have checked that we similar results for the other bins as well) , $\sigma(z^{(2)}_{\mbox{\scriptsize{b}}})$, for two samples with size  $N_{\mbox{\scriptsize{g}}} = 5 \times 10^7$ and photo-z scatter $\sigma_{z,0} = 0.02$ and $\sigma_{z,0} = 0.08$ respectively, when the cross-correlations (i) are included and (ii) are not included. We see that the inclusion of cross-correlations improves the self-calibration level of the photo-z parameters. The improvement is more significant for the sample with larger $\sigma_{z,0}$; although in the case without cross-correlations the sample with $\sigma_{z,0}=0.02$  gives tighter constraints on the photo-z parameters, the sample with $\sigma_{z,0}=0.08$ gives better constraints in the case with cross-correlations.

In this section we explored the significance of including the cross-correlations in cosmological analyses of angular galaxy clustering. We find that for a standard, wide ($\delta z = 0.15$) binning choice, larger samples with worse photo-z scatter result in a significantly higher FoM compared to smaller, more accurate samples ($\sim 10-20$  higher FoM for a reasonable range of priors). Note that here we did not perform a detailed study of the dependence of our results on the priors, as we did in Section \ref{subsec: Depend_on_priors}. From Fig. \ref{fig: With_cross}, however, we see that although the relative difference in the FoM changes with different priors, the general picture remains the same.

\section{Dependence on the bin size}
\label{sec: Bin_size}

\begin{figure*}
\centering
\subfigure[]{\includegraphics[width=\columnwidth]{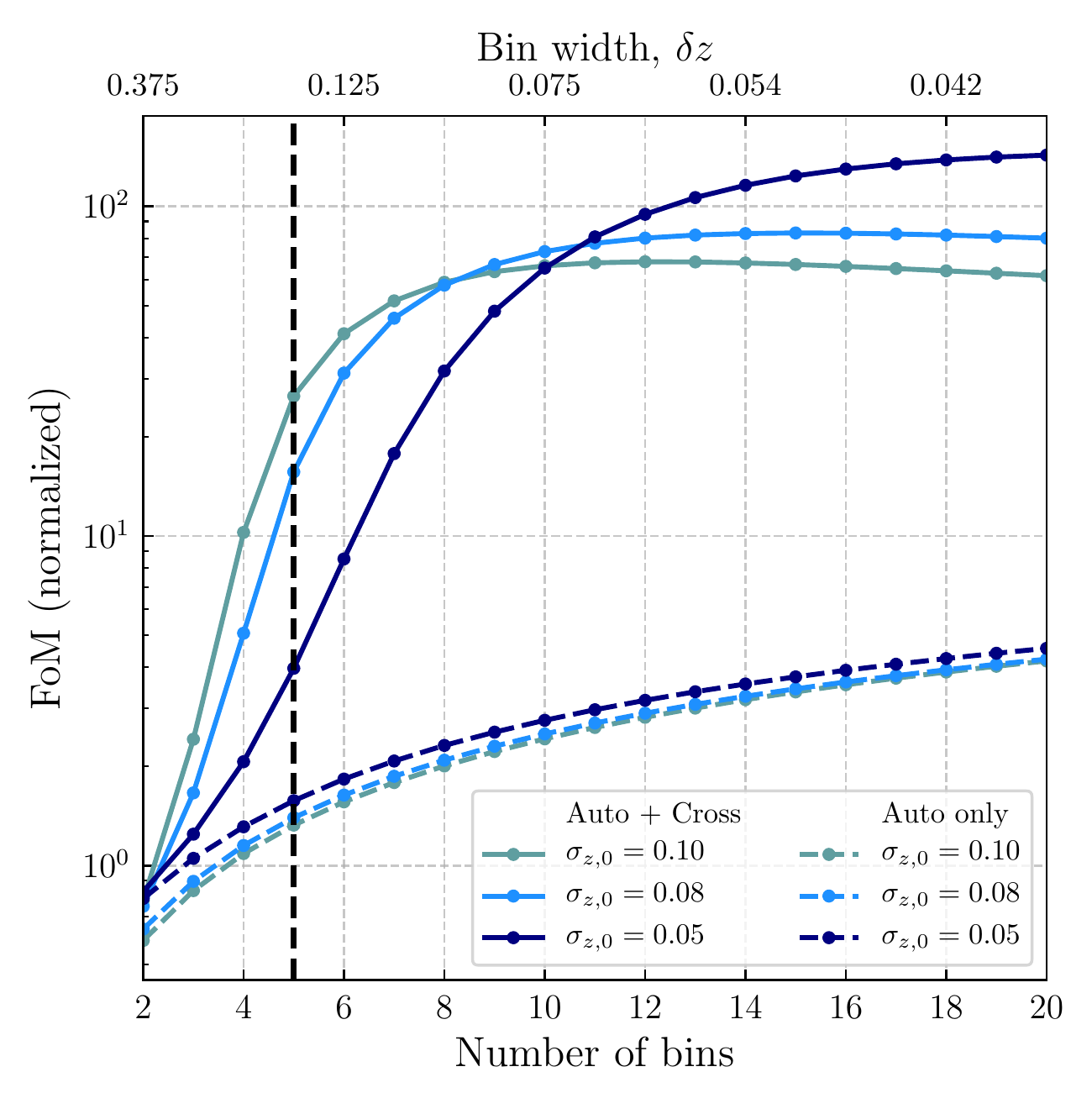}}
\subfigure[]{\includegraphics[width=\columnwidth]{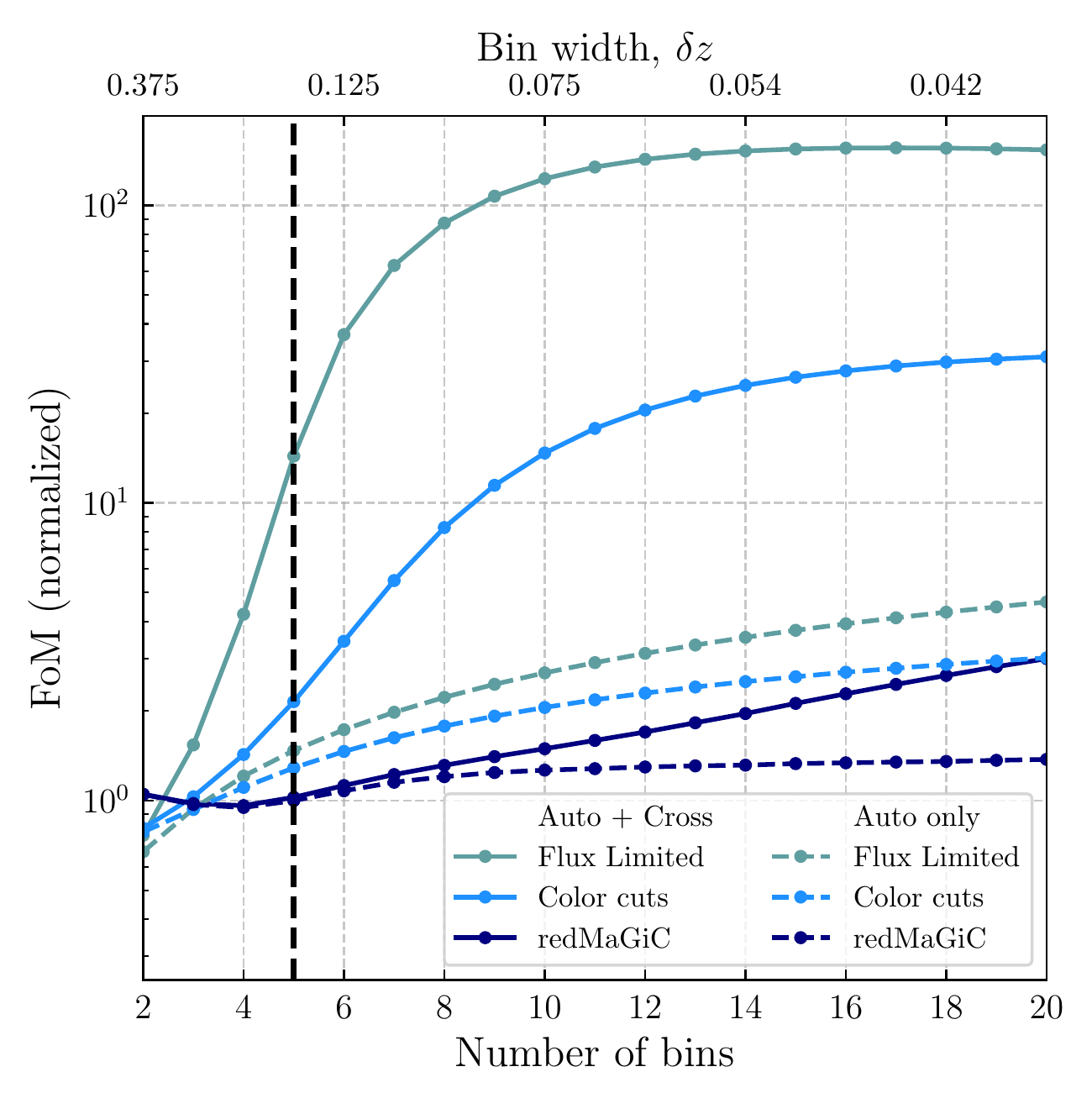}}
\caption{The dependence of the FoM on the number of bins used, or equivalently the bin width. We consider  the cases where only the auto-spectra are included (dashed lines) and both auto- and cross-spectra are taken into account (solid lines). In panel (a) we present the results for three samples all with size $N_{\mbox{\scriptsize{g}}} = 4 \times 10^7$ and different values for the photometric uncertainty scatter parameter. In panel (b) we present the results for the redMaGiC, and flux-limited  and color-cuts defined samples. In all cases we assume conservative priors of the form $\sigma(\sigma_{z,0}) = \sigma(z_{\mbox{\scriptsize{b}}}) = 0.4  \sigma_{z,0}$ on the photo-z parameters. We note  that cross-correlations become more and more important when the photo-z scatter is high compared to the bin size. We normalize to the FoM of the redMaGiC sample, from auto-correlations only, in the standard five bins used in the previous section. The vertical line corresponds to 5 bins, the baseline case we used in the previous sections.  }
\label{fig: Number_of_bins}
\end{figure*}

In this section we study in detail the dependence of our results on the bin width or, equivalently, the number of bins in the redshift range $z \in [0.20-0.95]$. 

In the panel (a) of Fig. \ref{fig: Number_of_bins} we plot the FoM for three samples with the same sample size ($N_{\mbox{\scriptsize{g}}} = 4 \times 10^7$) and different $\sigma_{z,0}$  (0.05, 0.08, 0.10) as a function of the number of redshift bins. We choose a range of two to twenty bins in the redshift range mentioned above. We present with dashed lines the case where only the auto-spectra are included in the calculation of the FoM and with solid lines the case where we consider both the auto- and cross-spectra. We use conservative photo-z priors ($\sigma(\sigma_{z,0}) = \sigma(z_{\mbox{\scriptsize{b}}}) = 0.4  \sigma_{z,0}$). We normalize everything to the FoM of the redMaGiC sample with auto-spectra only and five redshift bins.

We observe several interesting trends. In all cases the FoM is much higher (except in the limit of two redshift bins when all results seem to converge) when cross-correlations are included, in accordance to our findings in the previous section. The FoM from auto-correlations-only is almost the same for the three samples; see the flattening of the iso-FoM curves in panel (b) of Fig. \ref{fig: Baseline}. Furthermore, we see that as we increase the number of bins it continues to grow without showing a sign of saturation in the range of bins considered. 

In the case where both auto and cross spectra are considered the FoM initially grows very rapidly as we increase the number of bins and subsequently saturates when a large number of bins is considered. Note that the FoM of samples with higher $\sigma_{z,0}$ saturates at a lower number of redshift bins. The saturation is expected when the bin width $\delta z$ drops below the photo-z error, $\sigma_{z,0}$, since then there is no more information (see also \citealt{Asorey2012}) and this behavior is observed here. When the number of bins is low (less than $\sim 8-10$), samples with smaller $\sigma_{z,0}$ have higher FoM; cross-correlations carry more information for such samples. In the high bin number limit, on the other hand, samples with more smaller $\sigma_{z,0}$ have higher FoM. This explains the counter-intuitive results of the previous section: for samples with the same size, those with smaller value of $\sigma_{z,0}$ carry intrinsically more information and they can result to tighter cosmological constraints; but to do so we need a large number of bins that fully exploits their potential.

In panel (b) of Fig. \ref{fig: Number_of_bins} we present a similar plot for the RM, FL and CC samples. The FoM in the auto-spectra only case is not the same now; the FL samples seems to have a consistently higher FoM, increasing with the number of bins; the FoM of the RM is slightly increasing and almost saturates for a large enough number of bins.  As expected from the previous discussion, when cross-correlations are included the FoM of the FL sample grows rapidly with the number of bins but it soon saturates (for $\sim 10$ bins). Due to their smaller $\sigma_{z,0}$, the FoM  of the CC and RM samples continues to grow; especially for the RM sample, it seems to be far from saturation even when 20 bins in the range $z \in [0.2,0.95]$ are considered, and significantly lower than that of the FL (and CC) sample. This is of course expected, since even for 20 bins, the bin width is $\delta z = 0.0375$, which is larger than the value of the $\sigma_{z,0}$ for the RM sample. In practice, even for 40 bins (when $\delta z \sim \sigma_{z,0}$) the RM still gives a FoM that is significant smaller than that the FL gives, since the RM is much smaller in size.

The above discussion suggests that if we want to extract the maximum possible information from samples with accurate photo-zs a large number of redshift bins is required. This also points to the fact that, for such samples it may be worthy to consider a continuous formalism to analyze the 3D overdensity field, instead of considering (a large number of) tomographic bins. See, for example \cite{Lanusse2015,Passaglia2017}. For the binning choice of DES Y1 (5 bins) or the expected 10 bins in LSST, larger samples with worse photo-zs outperform samples like the RM because of the information carried in the cross spectra and the better self-calibration of the photo-z parameters.

\section{Summary}
\label{sec: Summary}

In this paper we have presented a systematic study of optimizing the galaxy sample selection in order to maximize cosmological constraints from galaxy clustering. We study the trade-offs between the photometric redshift uncertainty scatter, $\sigma_{z,0}$ , and galaxy sample size, $N_{\mbox{\scriptsize{g}}}$, under different analysis scenarios (redshift bin sizes, inclusion or not of cross-correlations, priors on the photo-z parameters).

We use a simple model to describe different galaxy samples in a unified way. We use the standard Fisher formalism to forecast cosmological constraints. The main assumptions of our approach are:

$\bullet$ We use a common underlying redshift distribution, of the form \eqref{eq: red_dist}, for all samples.

$\bullet$ We assume Gaussian photo-zs, characterized by a scatter parameter, $\sigma_{z,0}$, and one redshift bias parameter, $z_{\mbox{\scriptsize{b}}}$, per bin.

$\bullet$ We use different priors on the photo-z parameters, all of the form $\sigma(\sigma_z) = \sigma(z_b^i) \propto \sigma_z$.

$\bullet$ We use a common galaxy bias for all samples, of the form $b(z) = 1+ z$, that we keep fixed to its fiducial value.

As a primary example of a photometric survey we use the Dark Energy Survey (DES). The three specific samples used in this paper are: the redMaGiC (RM) sample (that was used in Y1 clustering analyses), a flux-limited (FL) sample and a red-galaxies-dominated sub-sample of the flux-limited sample, defined through color cuts (CC) (see \ref{subsec: Sample_Selection} for their definition, sizes, and photo-zs). We scale the sample sizes to match the expected DES  Y3 footprint. 

We note that the above model is a better approximation to the RM sample and, generally, samples dominated by red galaxies; for those samples the redshift uncertainties are well understood and can be described using a Gaussian distribution. For the FL sample (and generally large samples, dominated by blue galaxies) photo-zs cannot accurately modeled as following a simple Gaussian; for example, a model with a mixture of Gaussians -- and thus more photo-z parameters -- would be in practice more realistic.

We find that our forecasts of the constraints on the cosmological parameters $\Omega_m,\, \sigma_8$ using the simple model described above, with a common underlying redshift distribution and a common galaxy bias agree to a $\sim 3 \% - 10\%$ level to the results obtained using the redshift distributions and galaxy biases derived from the data, for clustering in five redshift bins between $z = 0.2$ and $z = 0.95$, considering auto correlations only. This suggests that the simplifying assumptions listed above lead to an uncertainty in the final result roughly at the $10 \%$ level.

We forecast the constraints on $\Omega_m, \sigma_8$ from angular clustering in the same five bins as before, using different samples in the photo-z error range $0.01 \leq \sigma_{z,0} \leq 0.1$ and size range $1.0 \times 10^6 \leq N_{\mbox{\scriptsize{g}}} \leq 8.0 \times 10^8$. Our main findings are the following:

$\bullet$ When we use only the auto-correlation spectra in our calculations, there seems to be a gain from using a large sample, with high $\sigma_{z,0}$ (like the flux-limited sample) instead of a smaller sample with accurate photo-zs (like the redMaGiC), of the order of $\sim 20 \% -  50 \%$. These conclusions hold for a range of plausible priors on the photo-z parameters.

$\bullet$ When cross-correlations are included in the analysis, our results show a completely different behavior compared to the auto-spectra-only case. Specifically, there is a very significant cosmological information gain when using large samples with high values of $\sigma_{z,0}$. For example, using the flux-limited sample, we get a figure of merit (FoM) that is $\sim 10 - 20$ times higher (for reasonable photo-z priors) than that we get from the redMaGiC sample. This can be explained by the fact that, for samples with high $\sigma_{z,0}$ the overlap between redshift bins is significant, and so is the information we get from the inclusion of the cross-correlations in the analysis. Furthermore, this overlap leads to a significant self-calibration of the photo-z parameters which, in turn, leads to better constraints on the cosmological parameters.

$\bullet$ We found that both the relative FoM from different samples, as well as the absolute value of the FoM are significantly affected by the level of the external calibration of the photo-z parameters, which is quantified by the external priors on them.

Finally, we considered the effect  of the redshift bin width, or equivalently the number of redshift bins in the $0.2 - 0.95$ redshift range, on our results. We find that (in the  case when cross-correlations are included), for a small number $(5-10)$ of bins, those samples with high $\sigma_{z,0}$ give a higher FoM, for the same reasons explained above. The FoM increases with the number of bins and saturates when the bin size is comparable to the photo-z scatter, $\delta z \sim \sigma_{z,0}$. For samples with accurate photo-zs (low $\sigma_{z,0}$) the FoM continues to grow even when a large number of bins is considered and finally a higher FoM is achieved, a result of the fact that these samples carry intrinsically more information, that can be retrieved with the appropriate thin binning. However, we note that a direct comparison (that the sample with smaller $\sigma_{z,0}$ will finally result a higher FoM) is possible only for samples of the same size; samples like the redMaGiC, even if we consider a very high number of bins ($\sim 40$, such as the bin width to be $\delta z \sim 0.017$) can still result in a lower FoM compared to much larger samples, like the flux-limited sample.

We conclude that especially for typical binning choices, where a number of $5-10$ redshift bins is considered, there can be a significant gain from using larger samples, despite their larger $\sigma_{z,0}$, if the cross-correlation spectra are included in the galaxy clustering analysis.

Although our analysis used DES as a primary example of a photometric galaxy survey, we expect our main qualitative results to hold for other surveys as well, like the upcoming LSST survey.

\section*{Acknowledgements}

We thank Martin Crocce, Scott Dodelson, Jack Elvin-Poole, Samuel Passaglia, Anna Porredon, and Eduardo Rozo for helpful suggestions and discussions. The confidence ellipses of Fig. \ref{fig: Data_and_model} were plotted using the CosmicFish package for cosmological forecasts \citep{Raveri2016}.







\appendix

\section{Results with fixed photo-z scatter}
\label{sec: shift_and_Gaussian}

\begin{figure*}
\centering
\subfigure[]{\includegraphics[width=\columnwidth]{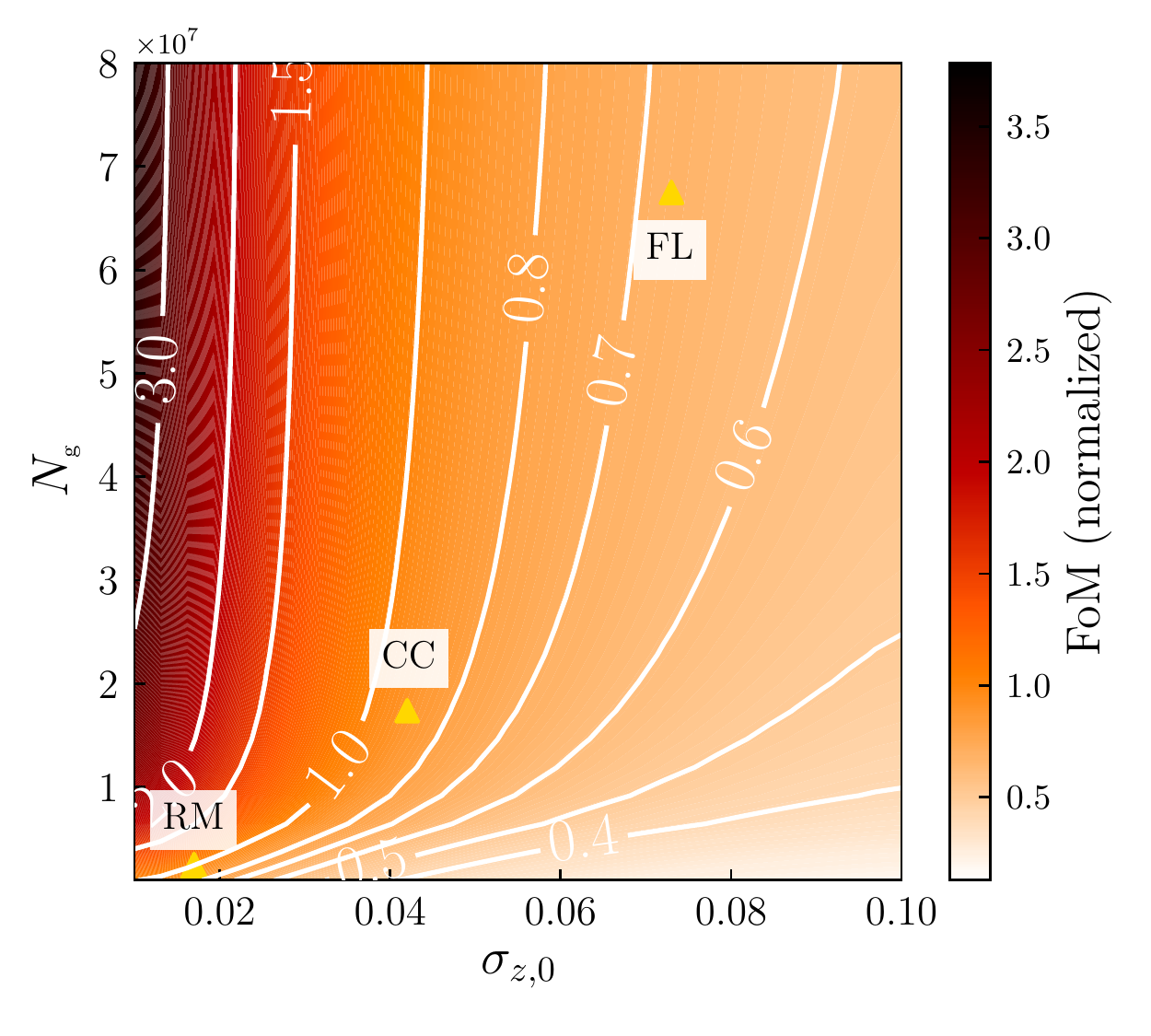}}
\subfigure[]{\includegraphics[width=\columnwidth]{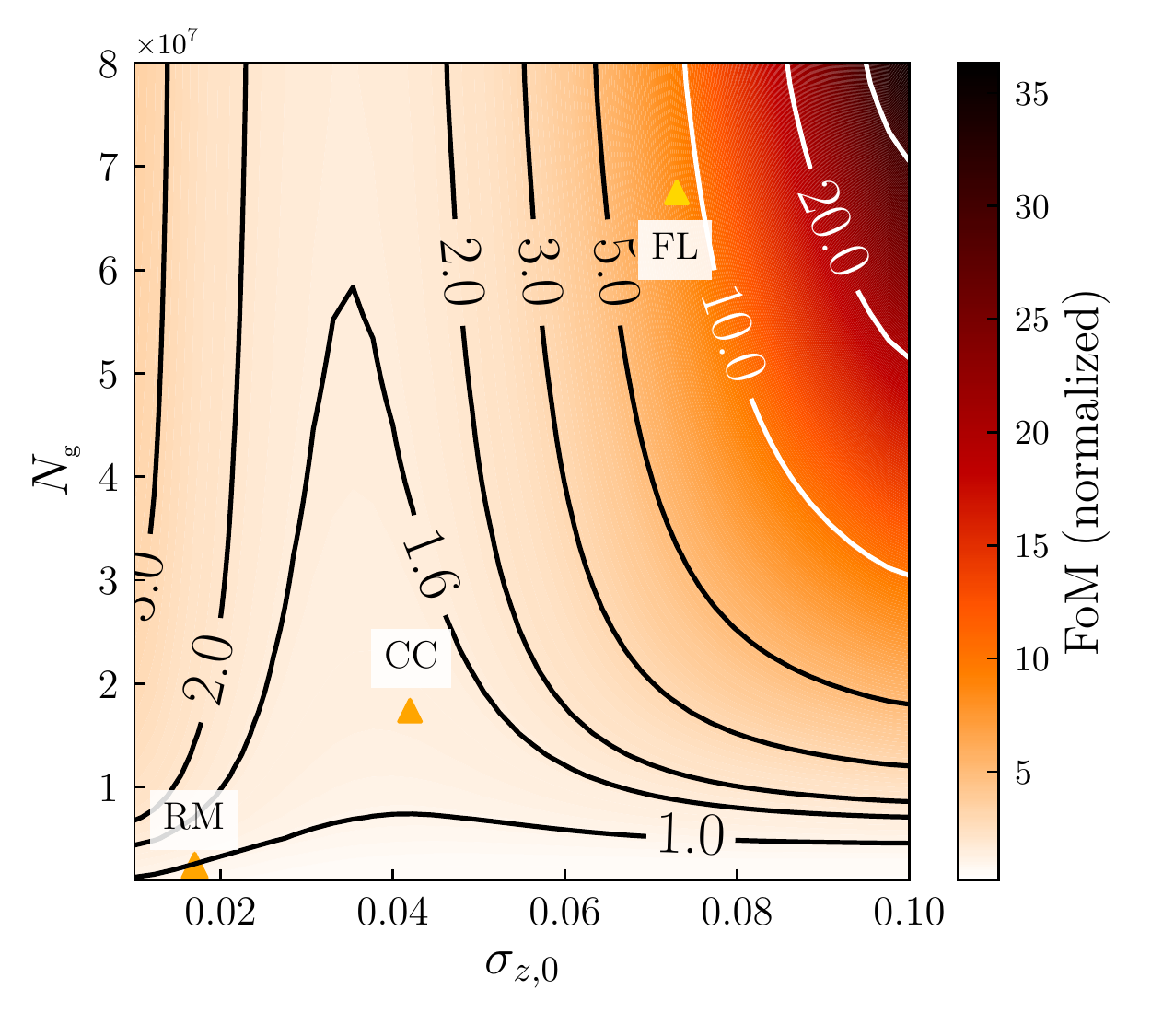}}
\caption{{\textit{Panel} (a)}: As in Fig. \ref{fig: Baseline} (no cross-correlations), panel (b), but now fixing the photo-z scatter parameter, $\sigma_{z,0}$, to its fiducial value for each sample. {\textit{Panel}} (b): As in Fig. \ref{fig: With_cross} (with cross-correlations), panel (b), now with fixed $\sigma_{z,0}$.}
\label{fig: Fixed_phot_z}
\end{figure*}

In the main text we performed our analysis marginalizing over the photo-z scatter parameter, $\sigma_{z,0}$ with priors of the order $\sigma(\sigma_{z,0}) = (0.4 - 0.04)\sigma_{z,0}$. However, in Fig. \ref{fig: Data_and_model} we saw that this choice gives different results compared to the case where $\sigma_{z,0}$ is kept fixed to its fiducial value, an approach that is closer to the treatment of photo-z uncertainty in DES analyses.

Here we briefly examine how some of our results change when we fix the photo-z scatter parameter and leave free only the photo-z bias parameters.

In Fig. \ref{fig: Fixed_phot_z} we present the  FoM  (normalized to the FoM of the redMaGiC sample) from angular clustering in the same five bins and for the same range in size and photo-z scatter parameter as we did in the main text, but now keeping $\sigma_{z,0}$ fixed. In panel (a) we show results using the auto-correlation spectra only, as in Section \ref{sec: Baseline}, while in panel (b) we include cross-correlations as well, as in Section \ref{sec: Cross_correlations}. In both cases we impose conservative priors  on the photo-z bias parameters, of the form $\sigma(z_{\mbox{\scriptsize{b}}}) = 0.4\sigma_{z,0}$ (thus compare Fig. \ref{fig: Fixed_phot_z} with panels (b) in Figs. \ref{fig: Baseline} and \ref{fig: With_cross}).

The relative FoM between samples is of the same order of magnitude as in the case presented in the main text in both cases. However there are some differences. For example, in the auto-spectra only case (panel (a)), the FL sample gives a lower FoM than the RM one.  In the case where cross-correlations are included (panel (b)) the FL still gives a significantly higher ($\sim 8.4$ times higher) FoM than the RM sample, but the difference is now significantly lower (in the main text it was $\sim 14 $ times higher).

Finally, note that when we fix the photo-z scatter parameter, the overall constraints significantly improve. The ratio of the FoM, for the redMaGiC sample, between the case discussed here and the case discussed in the main text is:

\begin{equation}
\frac{\mbox{FoM}_{\mbox{\scriptsize{RM}}}^{\mbox{\scriptsize{fixed}}}}{\mbox{FoM}_{\mbox{\scriptsize{RM}}}^{\mbox{\scriptsize{free}}}} \cong 4.4.
\end{equation}

\section{Derivatives of the Angular power spectra with respect to the photo-z parameters}

The derivatives of the angular auto- and cross-spectra, with respect to the photo-z parameters, that enter into the calculation of the Fisher matrix, can be calculated analytically. 
Here we present these analytic expressions, both for the case where photo-zs errors are modeled as following a Gaussian distribution with scatter and one photo-z bias per redshift bin, and for the case where photo-z uncertainties are introduced as shifts in the observationally obtained redshift distributions.

Let us start by calculating the derivative of the angular spectra with respect to the photometric redshift error spread, $\sigma_{z,0}$, in the Gaussian photo-z case. The dependence of the angular spectra on $\sigma_{z,0}$ comes only through the weighting kernel for angular clustering, $W^i(z)$, $i = 1, \dots,  N_{\mbox{\scriptsize{bins}}}$, so the derivative in this case can be written:
\begin{align}
\frac{\partial C_\ell^{ij}}{\partial \sigma_{z,0}}  &=  \int  dz \frac{H(z)}{c\chi^2(z)} \times \nonumber \\   
 & \left(\frac{\partial W^i(z)}{\partial \sigma_{z,0}} W^j(z) +
 \frac{\partial W^j(z)}{\partial \sigma_{z,0}} W^i(z)  \right)P_{NL}\left(k=\frac{\ell+1/2}{\chi(z)},z \right) 
\end{align}

From the definition of the clustering kernel, we have:
\begin{equation}
\frac{\partial W^i(z)}{\partial \sigma_{z,0}} =b_g^i\frac{dN_{\mbox{\scriptsize{g}}} }{dz}\frac{\frac{\partial F^i(z)}{\partial \sigma_{z,0}}\int \frac{dN_{\mbox{\scriptsize{g}}} }{dz'}F^i(z')dz' -F^i(z)\int \frac{dN_{\mbox{\scriptsize{g}}} }{dz'} \frac{\partial F^i(z')}{\partial \sigma_{z,0} }dz'}{\left(\int \frac{dN_{\mbox{\scriptsize{g}}} }{dz'}F^i(z')dz' \right)^2},
\end{equation}
with:
\begin{equation}
\frac{\partial  F^i(z)}{\partial \sigma_{z,0}} = \frac{1}{\sqrt{\pi}\sigma_{z,0}}\left[ x_{\mbox{\scriptsize{max}}}^i e^{- (x_{\mbox{\scriptsize{max}}}^i)^2}  -  x_{\mbox{\scriptsize{min}}}^i e^{- (x_{\mbox{\scriptsize{min}}}^i)^2} \right],
\end{equation}
where we used that $\frac{d}{dz} \mbox{erf}(z) =  \frac{2}{\sqrt{\pi}} e^{-z^2}$ and $x_{\mbox{\scriptsize{max}}}^i$ is defined in Eq. \eqref{eq: chi_def}.

For the photo-z biases, $z_b^i$, $i = 1, \dots, N_{\mbox{\scriptsize{bins}}}$ we get the derivatives:
\begin{align}
\frac{\partial C_\ell^{ij}}{\partial z_b^k}  &=  \int  dz \frac{H(z)}{c \chi^2(z)}  \times \nonumber \\   
 & \left(\delta^{ik}\frac{\partial  W^i(z)}{\partial z_b} W^j(z) +
 \delta^{jk}\frac{\partial W^j(z)}{\partial z_b} W^i(z)  \right)P_{NL}\left(k=\frac{\ell+1/2}{\chi(z)},z \right) 
\end{align}
The derivative of the weighting kernel is the same as before with the substitution $\frac{\partial F^i}{\partial \sigma_{z,0}} \to \frac{\partial F^i}{\partial z_b}$, where:
\begin{equation}
\frac{\partial F^i(z)}{\partial z_b}  = \frac{1}{\sqrt{2\pi}\sigma_{z,0}(1+z)} \left[e^{- (x_{\mbox{\scriptsize{max}}}^i)^2} - e^{- (x_{\mbox{\scriptsize{min}}}^i)^2} \right].
\end{equation}

In the case where instead of assuming a photo-z model, we get the galaxy  distributions at each redshift bin, $\hat{n}_g^i(z)$ by introducing one shift per bin, $\hat{n}_g^i(z-\Delta z^i)$. The derivatives of the angular spectra with respect to the shift $\Delta z^k$ will now be:

\begin{align}
&\frac{\partial C_\ell^{ij}}{\partial \Delta z^k}  =  \int  dz \frac{H(z)}{c}\frac{b_g^i b_g^j}{\chi^2(z)} \times \nonumber \\   
 & \left(\frac{\partial \hat{n}_g^i(z - \Delta z^i)}{\partial \Delta z^k}\hat{n}_g^j(z - \Delta z^j)  +
 \frac{\partial \hat{n}_g^j(z - \Delta z^j)}{\partial \Delta z^k} \hat{n}_g^i(z - \Delta z^i)   \right) \times \nonumber \\
 &P_{NL}\left(k=\frac{\ell+1/2}{\chi(z)},z \right), 
\end{align}

with
\begin{equation}
\frac{\partial \hat{n}_g^i}{\partial \Delta z^i} = - \delta^{ik} \frac{\partial \hat{n}_g^i(x)}{\partial x},
\end{equation}
where $x = z - \Delta z^i$.


\bsp	
\label{lastpage}
\end{document}